\documentclass[a4paper]{iopart}
\usepackage{iopams}  
\usepackage[latin1]{inputenc}
\usepackage{bm}
\usepackage{BibDef}
\usepackage{latexsym}
\usepackage{color}
\usepackage{graphicx}
\usepackage{caption}
\usepackage{cite}
\usepackage{url}

\usepackage{hyperref}
\hypersetup{
  colorlinks=true,        
  linkcolor=blue,         
  citecolor=cyan,         
}

\def\lsim{\mathrel{\rlap{\lower 3pt \hbox{$\sim$}} \raise 2.0pt \hbox{$<$}}}
\def\gsim{\mathrel{\rlap{\lower 3pt \hbox{$\sim$}} \raise 2.0pt \hbox{$>$}}}

\def\ud{{\rm d}}

\begin{document}

\setcounter{footnote}{0}

\title[About gravitational-wave generation by a three-body system]{About
  gravitational-wave generation 
  by a three-body system}

\author{Matteo Bonetti$^{1,2}$, Enrico Barausse$^{3}$, Guillaume Faye$^{3}$,
  Francesco Haardt$^{1,2}$ \& Alberto Sesana$^{4}$}

\address{$^1$DiSAT, Universit\`a degli Studi dell'Insubria, Via Valleggio 11,
  22100 Como, Italy\\
$^2$INFN, Sezione di Milano-Bicocca, Piazza della Scienza 3, 20126 Milano, Italy\\ 
$^3$ $\mathcal{G}\mathbb{R}\varepsilon{\mathbb{C}}\mathcal{O}$,
  Institut d'Astrophysique de Paris, UMR 7095, CNRS,\\ Sorbonne
  Universit{\'e}s \& UPMC Univ Paris 6, 98\textsuperscript{bis}
  boulevard Arago, 75014 Paris, France\\
  $^4$School of Physics and Astronomy, University of
Birmingham, Edgbaston, Birmingham B15 2TT, United Kingdom\\}
\ead{mbonetti@studenti.uninsubria.it}
\vspace{10pt}

\begin{abstract}
  We highlight some subtleties that affect naive implementations of
  quadrupolar and octupolar gravitational waveforms from
  numerically-integrated trajectories of three-body systems. Some of
      those subtleties arise from the requirement that the source be contained
    in its ``coordinate near zone'' when applying the standard PN
    formulae for gravitational-wave emission, and from the need to use the
    non-linear Einstein equations to correctly derive the quadrupole emission
    formula. We show that some of these subtleties were occasionally
  overlooked in the literature, with consequences for published results. We
  also provide prescriptions that lead to correct and robust predictions for
  the waveforms computed from numerically-integrated orbits.
\end{abstract}

%
%
%
%
%

\section{Introduction}

While the two-body problem is completely solvable in Newtonian theory, no
general exact solution to it is known in General Relativity (GR). As a result,
the dynamics of a binary system in GR can only be obtained by solving
perturbatively the field equations, or through numerical techniques on a
computer (``numerical
relativity''~\cite{Pretorius2005,Campanelli2005,Baker2005,
    Alcubierre2008_book,Lehner2014}). Perturbative schemes, valid in
different regimes, include the Post-Newtonian (PN)
approximation~\cite{Blanchet2014}, which consists in expanding the dynamics in
powers of $v/c \ll 1$\footnote{In the standard PN book-keeping a term
    suppressed by a factor $(v/c)^{2n}$ with respect to the leading (i.e.,
    Newtonian) order is said to be of $n$PN order.} ($v$ being the relative
velocity of the binary, $c$ the speed of light in vacuum) and the self-force
formalism, which instead relies upon an expansion in the binary's mass ratio,
assumed to be small~\cite{Poisson2004}. The detection of gravitational waves
(GWs), indirectly from binary pulsars~\cite{Hulse1974} and directly from
systems of two merging black holes
(BHs)~\cite{GW150914,GW170104,TheLIGOScientific2016pea}, provides an excellent
benchmark to test the general-relativistic two-body dynamics. As a matter of
fact, these observations are to date in perfect agreement with the GR
predictions~\cite{Taylor1989,Taylor1982,Damour1992,TheLIGOScientific2016src,
  TheLIGOScientific2016pea,GW170104}.

Unsurprisingly, the three-body problem, which already in Newtonian theory does
not admit any general closed form solution and, moreover, gives rise to
chaotic dynamics, becomes considerably harder in GR. However, its
general-relativistic dynamics can be obtained within the PN approximation scheme,
just like in the two-body case. Indeed, one can write a PN-expanded,
time-dependent
Hamiltonian~\cite{Schafer1987,Konigsdorffer2003,Lousto2008,Galaviz2011} that
describes the conservative dynamics of a system of three non-spinning bodies
up to the 2PN order (i.e., through order $(v/c)^4$ beyond the leading order
Newtonian dynamics), as well as its dissipative dynamics (i.e., the
back-reaction due to GW emission) to leading order in $v/c$, which corresponds
to a contribution of 2.5PN order, or ${\cal O}(v/c)^5$, in the equations of
motion.

The PN dynamics of three body systems is not just an academic curiosity.
Kozai-Lidov resonances~\cite{Kozai1962,Lidov1962,Ford2000,Naoz2016}, first
discovered in Newtonian triplets, are believed to be a relevant astrophysical
mechanism for the formation of binary systems of stellar-mass BHs (observable
by ground-based GW detectors) in dense stellar
environments~\cite{Antonini2016,Antonini2017} or in
  isolation~\cite{Silsbee2017}. It may also play an important role in the
formation and evolution of binaries of supermassive
BHs~\cite{Blaes2002,Iwasawa2006,Iwasawa2008,Hoffman2007,Pau2010}, whose GW
signal is targeted by existing pulsar-timing
arrays~\cite{Haehnelt1994,Jaffe2003,Wyithe2003,Enoki2004,Sesana2004,Sesana2005,
  Jenet2005,Rhook2005, Barausse2012,Klein2016} and by the future space-borne
interferometer LISA~\cite{Audley2017}. It turns out that the PN corrections
are crucial to assess the efficiency of the Kozai-Lidov mechanism, as they can
destroy the resonance on which it relies. Indeed, the coherent piling up of
the perturbation induced by the third body may be disrupted due to
relativistic precession effects appearing at 1PN order and
beyond~\cite{Holman1997,Ford2000,Blaes2002}. Finally, the GW emission from
systems of three BHs has been studied in detail by means of numerical
  techniques~\cite{Campanelli2006,Campanelli2007,Lousto2007,Pau2010,
Galaviz2011}
in the event that they form and radiate in sufficiently large number to
provide a sizable population for GW detectors. For particular
  configurations, they have also been investigated
  analytically~\cite{Asada2009,Yamada2016} so as to gain some insight on their dynamics.

The GW emission from binary systems with relative velocities $v\ll c$ can be
modeled, at leading order, through the Einstein quadrupole
formula~\cite{Einstein1918a,Landau_book1975,Blanchet2014}. Next-to-leading
order corrections are given by the mass-octupole and current-quadrupole
contributions~\cite{Thorne1980,Blanchet2014}. A key requirement implicit in
the derivation of the corresponding formulae is that the binary must be
contained in its ``Near Coordinate Zone'' (NCZ), i.e., a region (centered on
the origin of the coordinates) of radius comparable to (but smaller than) the
GW wavelength $\lambda$. This requirement comes about because the PN
  formalism for GW generation, which can only be legitimately applied as
    long as the source is much smaller than $\lambda$, is based on a systematic
    multipole expansion of the gravitational field outside the source. In
    order to ensure an overlap between the domain of validity of this
    expansion (say $|\bm{x}|\gtrsim r_\mathrm{min}$) and the near zone (where the dynamics of the source is computed neglecting retardation effects), one
    must clearly have $r_\mathrm{min}\sim$ [size of the near zone]
    $\sim \lambda$, so  the coordinate origin and the source cannot be
    more than one wavelength apart\footnote{Nevertheless, the exact choice
    of where the NCZ is centered is a matter of definition. The important
    point is that it must contain both the whole source and the origin of the
    coordinates. In fact, one may alternatively think in terms of the binary's
    near zone, which is defined to be (roughly) centered on the center of mass
    (CoM) of the binary. In that case, a proper derivation of the quadrupole
    formula would require choosing the origin within the near zone. The
    adoption of this point of view would not alter any of the discussions of
    this paper.}. Indeed, these formulae are usually applied in the reference
  frame of the binary's CoM. In that frame, in the PN regime, the existence of
  a NCZ containing the binary is guaranteed, since the size of the system ---
  its separation $a$ --- is negligible relative to the wavelength
  $\lambda\sim a/(v/c)$.

\begin{center}
	\begin{figure}
		\centering
		\includegraphics[scale=0.5]{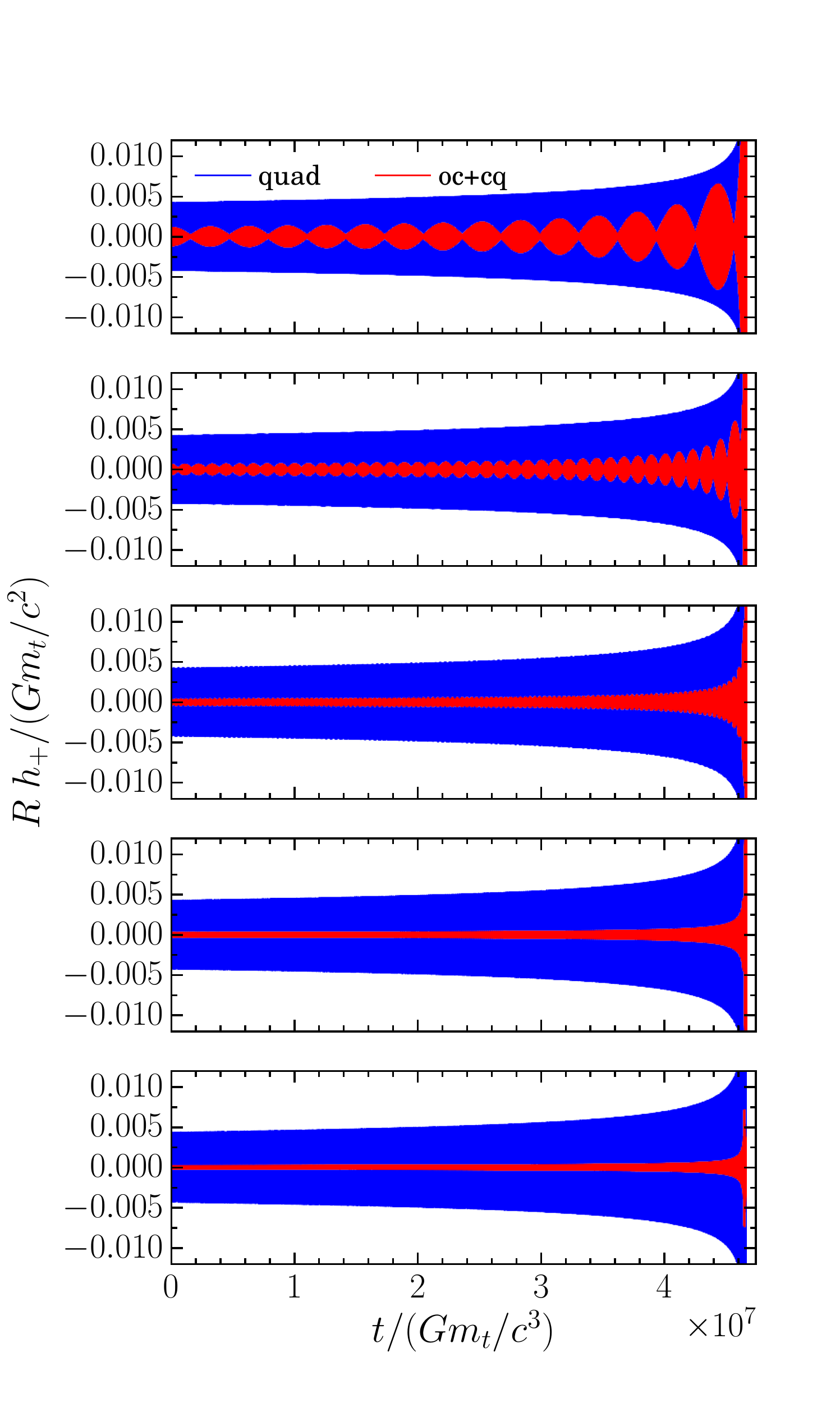}
		\caption{Waveforms from five triple systems with relative inclination
          $i=0$, inner separation $a_{\rm in} = 150 Gm_t/c^2$, inner
          eccentricity $e_{\rm in} = 0$, outer eccentricity $e_{\rm out} = 0$,
          and mass ratios $m_2/m_1 = 0.5$ and $m_3/(m_1+m_2) = 0.05$. From top
          to bottom $a_{\rm out} = [10000, 5000, 2500, 1250, 625]Gm_t/c^2$. The
          observer is located in the $xz$ plane of a fixed spatial frame ($x$,
          $y$, $z$), with spherical coordinates $\theta = \pi/4$, $\phi=0$. To
          be compared with~\cite{Galaviz2011}, fig.~18. As
          in~\cite{Galaviz2011}, the orbits are obtained by integrating
          numerically the Hamilton equations for the triple systems, through the
          2PN order in the conservative dynamics and at the leading (Newtonian)
          order in the dissipative one.}
		\label{fig:oct_triple0}
		\end{figure}
\end{center}

For a triple system with relative velocities $v \ll c$, it would seem natural
to apply the very same formulae in the reference frame of the CoM of the
three-body system. However, by doing so, one obtains unphysical results such as
those reported in fig.~18 of~\cite{Galaviz2011}, as we will now explain.
Indeed, we have reproduced the same behavior by applying the quadrupole and
``quadrupole-octupole'' formulae in the CoM reference frame of a series of triple
systems with mass ratios $m_2/m_1=0.5$ and $m_3/(m_1+m_2)=0.05$, whose
trajectories are computed with the code of~\cite{Bonetti2016} (which includes
the 1PN and 2PN conservative triple dynamics, and the leading order
dissipative dynamics). The ``inner binary'' (comprised of $m_1$ and $m_2$) of these
hierarchical triplets has zero initial
eccentricity and an initial separation $a_{\rm in}=150 Gm_t/c^2$, where $G$ is
Newton's constant and $m_t$ the total mass of the triplet (throughout the paper we instead reserve the symbol $m$ to indicate the total mass of binary systems, i.e., $m=m_1+m_2$). The ``outer binary'' (comprised of $m_3$ and the CoM of the inner
binary) has instead initial separation varying in the range
$a_{\rm out} \in [625,10000] Gm_t/c^2$, and zero initial eccentricity. The
results are displayed in fig.~\ref{fig:oct_triple0}, where one can observe,
paradoxically, that the effect induced by the third body
grows as it
gets farther away from the inner binary. We will analyze this situation in
detail in this paper, and show that the problem is connected to the fact that a
NCZ region centered on the CoM of the triplet and having size comparable to
the minimum gravitational wavelength excited by the system does {\it not}
include the whole triplet, unlike what happens for a binary system.

This, however, is just one example of the subtleties one should be aware of when
computing GW emission from binary or triple systems in a too naive
fashion. Another interesting apparent paradox arises, e.g., if one tries to
compute the gravitational waveforms of a binary (or triple) system by directly
integrating the equations for the linear perturbations $h_{\mu\nu}$ over a
background Minkowski space-time (endowed with a flat metric
$\eta_{\mu\nu}=\textrm{diag} (-1,1,1,1)$ and coordinates
$\{x^\mu\}_{\mu=0,1,2,3}$).

In the harmonic gauge, which is defined by the condition
$\partial_\mu \bar{h}^{\mu\nu}=0$, where $\partial_\mu$ is the flat
four-dimensional derivative and
$\bar{h}^{\mu\nu} = h^{\mu\nu} - 1/2 \ \eta^{\mu\nu}
h^{\alpha}_{\phantom{a}\alpha}$
represents the trace-reversed metric perturbation\footnote{In our conventions,
  space-time Greek indices are raised or lowered with the metric
  $\eta_{\mu\nu}$ or its inverse $\eta^{\mu\nu}$, whereas space Latin indices
  are raised or lowered with the Euclidean metric $\delta_{ij}$ or its inverse
  $\delta^{ij}$. In particular:
  $h^{\mu\nu} = \eta^{\mu\alpha} \eta^{\nu\beta} h_{\alpha\beta}$ and
  $h^{\alpha}_{\phantom{a} \alpha}= \eta^{\alpha\beta} h_{\alpha\beta}$.}, the
linearized Einstein equations read (see, e.g.,~\cite{Maggiore2007_book} \S 1.1,
  \cite{MTW} \S 35.1)
%
\begin{equation}\label{eq:first_wave_eq}
\Box_{\rm flat} \bar{h}^{\mu\nu}= -\frac{16 \pi G}{c^4} T^{\mu\nu}\,,
\end{equation}
%
where the d'Alembert operator
$\Box_{\rm flat}= \eta^{\mu\nu}\partial_\mu\partial_\nu$ is computed with the
background Minkowski metric and $T^{\mu\nu}$ is the source stress-energy
tensor.
These equations can be integrated exactly by using the (retarded) Green
function of $\Box_{\rm flat}$. The resulting waveforms (obtained from the
transverse trace-free part of the spatial components) may then be compared to those
predicted by the quadrupole formula (and its higher-order corrections that we
have mentioned above). 

\begin{center}
	\begin{figure}
	\centering
		\includegraphics[scale=0.45]{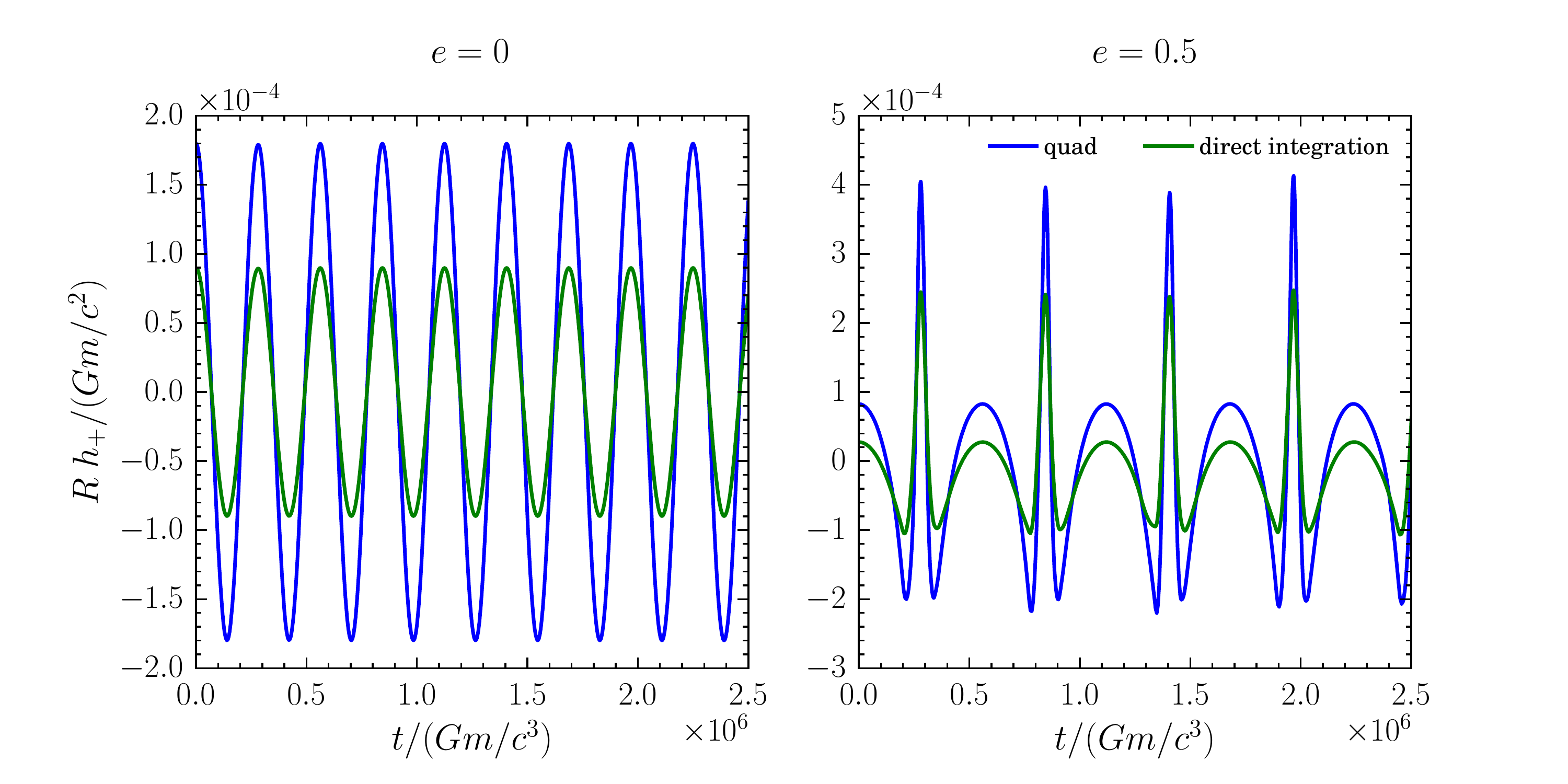}
		\caption{Quadrupole waveforms from two simulations of circular
          binaries with masses $m_1 = 0.9 m$, $m_2 = 0.1 m$. Blue lines are obtained with the quadrupole
          formula (see eq.~\ref{eq:first_quad_formula}); green lines are
          computed by direct integration of eq.~\ref{eq:first_wave_eq}. {\it
            Left panel}: circular case. {\it Right panel}: $e=0.5$.}
		\label{fig:quad_vs_green}
	\end{figure}
\end{center}

The comparison between the GW amplitudes obtained with the two
procedures for various binaries is shown in fig.~\ref{fig:quad_vs_green}. As
can be seen, there appears to be a factor $\sim 2$ discrepancy (this factor
becomes exactly 2 for binary circular orbits). Similar discrepancies arise
when integrating eq.~\ref{eq:first_wave_eq} for triple systems. This puzzling
difference will be discussed in more details. It is related to the fact, often
mentioned but rarely illustrated in introductory GR textbooks (see
however~\cite{Maggiore2007_book,MTW}), that a naive derivation of the
quadrupole formula based on eq.~\ref{eq:first_wave_eq} is wrong. It is
because that equation (via the harmonic gauge condition) implies that
$\partial_\mu T^{\mu\nu}=0$, which is clearly not verified for a binary system
since it entails that bodies move along straight lines.

The focus of this paper is thus pedagogical. We will discuss the two problems
mentioned above as well as other subtleties that we have encountered when
computing gravitational waveforms from numerically-integrated orbits of triple
systems. More precisely, the organization is as follows: in
Section~\ref{sec:cm_issue} we will illustrate, tackle and solve the problems
that arise when applying the standard quadrupole formula (and its higher-order
corrections) to a triple system. This will provide a solution to the
  discrepancy demonstrated in fig.~\ref{fig:oct_triple0}, which, as already
  mentioned, will turn out to be due to the source not being contained in its
  NCZ when the latter is centered on the triplet's center of mass. In
  Section~\ref{sec:green_solution} we will further comment about the
  inconsistent derivation of eq.~\ref{eq:first_wave_eq}, highlighting the need
  to use the \textit{non-linear} Einstein equations to compute GW emission
  self-consistently. Finally, in Section \ref{sec:conclusions} we will draw
our conclusions. Throughout the paper we use the $(-,+,+,+)$ signature
  convention.

\section{Emission of gravitational waves in hierarchical triplets}
\label{sec:cm_issue}

\setcounter{footnote}{0}

The leading-order contribution to the GW signal observed at space position
$\bm{x}$ and time $t$ is given, in an appropriate ``radiative'' gauge, by the
quadrupole formula (see~\cite{MTW} \S 36.10, \cite{Maggiore2007_book} \S
  3.3, \cite{Blanchet2014} \S 2.5)
%
\begin{equation}\label{eq:first_quad_formula}
h_{ij}^{\rm TT}(t,\bm{x}) = \frac{2 G}{R c^4} \Lambda_{ijkl}(\bm{n}) \ \frac{\ud^2
  M^{kl}(t_{\rm ret})}{\ud t^2} + \mathcal{O}\Big(\frac{1}{c^5}\Big)\, ,
\end{equation}
where $R=|\bm{x}|\equiv \sqrt{x^i x_i}$ is the distance of the observer
(assumed to be very far from the source compared with the wavelength $\lambda$ of the
emitted GWs),
$t_{\rm ret}=t-R/c$ is the retarded time of the background
space-time\footnote{Note that $t$ as appearing in
  eq.~\ref{eq:first_quad_formula} should rigorously be replaced by the
  radiative time $T = t - 2G M/c^3 \ln[R/(cb)]$, with $M$ being the total
  Arnowitt-Deser-Misner energy-mass~\cite{ADM1959} and $b$ representing some reference time.
  It is crucial to do so at future radiative infinity. However, since
  $2G M/c^3 \ln[R/(cb)]\ll R/c$, we may write $T \approx t$ for sufficiently
  large $R$ if $t$ remains bounded. See~\cite{Blanchet2002} for more details,
  in particular on how the logarithmic term is connected to the tail
  contribution to the waveform.}, $\Lambda_{ijkl}(\bm{n})$ denotes the
projector on the transverse-traceless (TT) gauge (see \ref{sec:app_A} for the
explicit definition), while
%
\begin{equation}\label{Q}
M^{ij}(t) = \int {\ud^3\bm{x}' \ c^{-2}T^{00}(t,\bm{x}') \Big(
  x'^ix'^j-\frac{1}{3}\delta^{ij}x'^kx'_k \Big)}
\end{equation}
represents the mass quadrupole moment of the source.
As mentioned in the introduction, implicit in the derivation of the quadrupole
formula (eq.~\ref{eq:first_quad_formula}) is the assumption that the source be
contained in its NCZ (see, e.g.,~\cite{MTW,Press1977,Maggiore2007_book}),
i.e., the reference frame in which the quadrupole moment (eq.~\ref{Q}) is
evaluated must be such that the source be contained within a region of size
$\sim \lambda$ centered on the origin of the coordinates. Finding a frame
satisfying this property is always possible for slowly moving binary systems,
since $\lambda$ is related to the system's typical (relative) velocity $v$ and
its typical separation $a$ by $\lambda\sim a/(v/c)$.

\begin{center}
	\begin{figure}[b]
		\centering
		\includegraphics[scale=0.42]{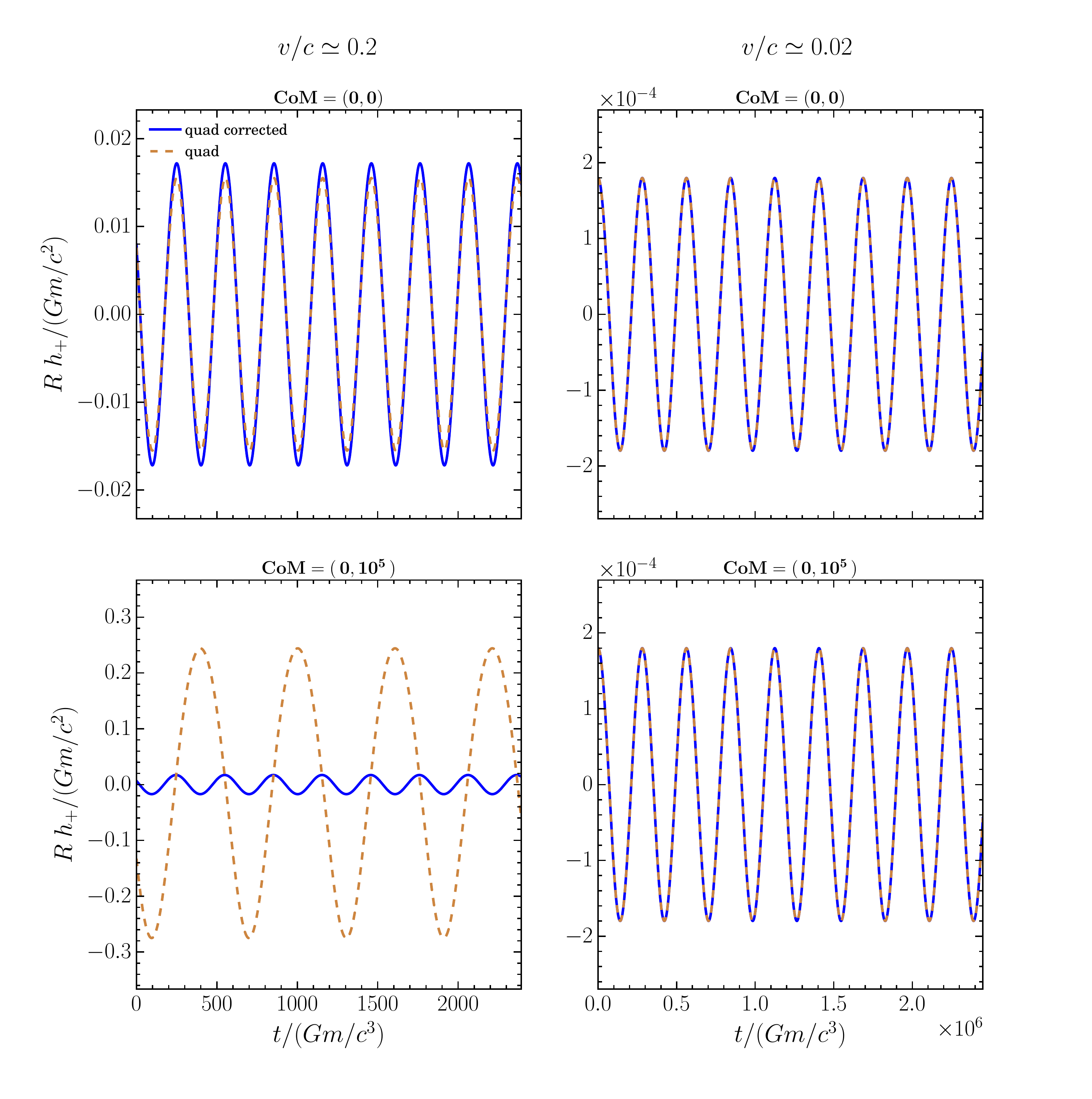}
		\caption{Quadrupole waveforms from four simulations of circular
          binaries with masses $m_1 = 0.9 m$, $m_2 = 0.1 m$ evolving
          according to the 1PN dynamics. {\it Left panels}: highly
          relativistic regime $v/c \simeq 0.2$. {\it Right panels}: mildly
          relativistic regime $v/c \simeq 0.02$. {\it Upper panels}: The
          binary's center of mass is placed in the origin of coordinates. {\it
            Lower panels}: The binary's center of mass is located at distance
          $10^5\times G m/c^2$ from the origin. The dashed lines represent
          quadrupole waveforms computed by simply inserting the trajectories
          of our simulations in eq.~\ref{eq:first_quad_formula}, while the
          solid blue lines are waveforms obtained from an ``amended''
          quadrupole formula (see text for details). The ``standard''
          quadrupole formula fails in the most relativistic and shifted binary
          case, whereas the amended one provides the correct result in all
          cases (note the different $y$-axis scales in the two left panels).}
	\label{fig:quad_corrected}
	\end{figure}
\end{center}

The most natural reference frame to describe the dynamics of an $N$-body
system is that of the CoM, where the equations of motion take their simplest
form. (Note that the usual Newtonian expression of the CoM position in terms
of the body locations $\bm{x}_1$, $\bm{x}_2$ for a binary, namely
$\bm{r}_0=(m_1 \bm{x}_1 + m_2 \bm{x}_2)/m$, is modified beyond the leading
order\footnote{The fact that the time derivative of the CoM position
      must be constant implies that the latter variable must be constructed in
      relation to a Noetherian current. In special relativity, this current,
      $\bm{r}_0-t \dot{\bm{r}}_0$, is nothing but the conserved quantity associated
      with the invariance of the dynamics under Lorentz boosts. Thus, the usual
      Newtonian definition has to be modified, the masses being
    replaced, notably, by the total energies of the bodies (see,
    e.g., \cite{Landau_book1975}, Vol II, \S 14 ). A similar extension of
      the Newtonian concept of CoM applies in GR
      (see~\cite{Landau_book1975}, Vol II, \S 96). We refer
    to~\cite{DS85,Georg2015} for an explicit construction in the case of
    binary systems.}.) In the case
under consideration here (i.e., a three-body system), a particularly
interesting configuration is that of the so-called hierarchical triplet. The
latter is comprised of an inner close binary $(m_1,m_2)$, supplemented by a
third body $m_3$ at larger distance. In practice, the system in this
configuration can be regarded as formed by two separate binaries: an inner
binary on the one hand, and an outer binary, formed by the third outer body
and the CoM of the inner binary, on the other hand.
From the point of view of the dynamics, the choice of the reference frame is
of course irrelevant. Indeed, the Hamiltonian (both at the Newtonian order and
when including the PN corrections) depends only upon the relative separations
of the three bodies, hence the dynamics of the system is frame-independent
(see e.g.,~\cite{Goldstein1950}). However, caution must be exercised when
applying the quadrupole formula (eq.~\ref{eq:first_quad_formula}) (and its
higher-order generalizations including octupolar corrections, etc.) to the
orbits resulting from numerical integrations of the equations of motion (see
fig.~\ref{fig:oct_triple0} and related discussion).

To illustrate this point, let us consider a circular binary located ``far'' from
the origin of the coordinates. In triple systems, this
happens for the inner binary when $m_3 a_{\rm out} \gg (m_1+m_2+m_3) a_{\rm in}$ since,
in this case, the inner binary is located far away from the CoM {\it of the triplet}
chosen as the origin of the coordinates. Setting the GW source in the $xy$
plane and the observer along the $z$ axis, eq.~\ref{eq:first_quad_formula}
takes the simplest possible form, i.e.,
%
\begin{equation}\label{eq:matrix_pol}
\fl h_{ij}^{\rm TT}(t,\bm{x}) = \frac{2 G}{R c^4} \left(
\begin{array}{ccc}
\frac{1}{2}[\ddot{M}_{11}(t_{\rm ret}) - \ddot{M}_{22}(t_{\rm ret})] 
  &
    \ddot{M}_{12}(t_{\rm ret})  & 0 \\
  \ddot{M}_{12}(t_{\rm ret})                     
  & -\frac{1}{2}[\ddot{M}_{11}(t_{\rm ret}) - \ddot{M}_{22}(t_{\rm ret})] & 0 \\
  0                                 & 0                                  & 0
\end{array}\right)\, ,
\end{equation}
where the dots placed over the quadrupole components $M_{ij}$ represent time
derivatives. The two independent polarizations of a propagating GW, referred to as
the ``plus'' and ``cross'' polarizations, are (in the situation considered here)
 simply the diagonal and off-diagonal part of eq.~\ref{eq:matrix_pol}:
%
\begin{eqnarray}
(h_+)_{\rm quad} &= \frac{G}{R c^4}(\ddot{M}_{11} - \ddot{M}_{22}),\nonumber\\
(h_{\times})_{\rm quad} &= \frac{2 G}{R c^4} \ddot{M}_{12}\, , \label{eq:matrix_pol2}
\end{eqnarray}
where all $M_{ij}$ components are evaluated at retarded time. Explicitly, for
a binary these expressions become (see, e.g., Problem 3.2
  of~\cite{Maggiore2007_book})
%
\begin{eqnarray}\label{eq:h_circ_bin}
(h_+)_{\rm quad} &= \frac{4G^2 m_1 m_2}{a R c^4}\cos(2\omega t_{\rm ret}),\nonumber\\
(h_{\times})_{\rm quad} &= \frac{4G^2 m_1 m_2}{ a R c^4}\sin(2\omega t_{\rm
                          ret})\, ,
\end{eqnarray}
$a$ denoting the separation of the binary and $\omega$ its orbital frequency. 

Let us now consider two different circular binaries, both with $m_1 = 0.9m$
and $m_2 = 0.1m$ but representative of two
different regimes: a rather relativistic binary with separation $a = 20 Gm/c^2$,
which corresponds to a relative orbital velocity $v/c \simeq 0.2$, and a
mildly-relativistic one, with $a = 2000 Gm/c^2$ (corresponding to
$v/c\simeq 0.02$). For these two systems, eqs.~\ref{eq:h_circ_bin} gives
$Rc^2/(Gm)(h_+)_{\rm quad} \simeq 1.8\times 10^{-2}$ and
$Rc^2/(Gm)(h_+)_{\rm quad} \simeq 1.8\times 10^{-4}$, respectively.
We then evolve them numerically in two different frames:
(1) one with the origin coinciding with the CoM, and (2) one with the origin
shifted by $10^5$ gravitational radii (i.e., $10^5\times G m/c^2$) from the CoM. Next,
we compute the waveforms directly via eqs.~\ref{eq:matrix_pol2} from the
numerical trajectories. Results are reported in fig.~\ref{fig:quad_corrected}
as dashed lines. The mildly-relativistic case is
consistent with the analytic predictions of eqs.~\ref{eq:h_circ_bin}. 
Instead, for the relativistic binary, $(h_{+})_{\rm quad}$ given by
eqs.~\ref{eq:matrix_pol2} is more than one order of magnitude higher than the
prediction from eqs.~\ref{eq:h_circ_bin} {\it when the origin of coordinates is
  far away from the CoM}. Indeed, we have checked that applying
eqs.~\ref{eq:matrix_pol2} directly to numerically-integrated trajectories
yields results that are coordinate-dependent. The discrepancy with
eqs.~\ref{eq:h_circ_bin} grows with the binary's relative velocity.

\begin{center}
	\begin{figure}[b]
		\centering
		\includegraphics[scale=0.42]{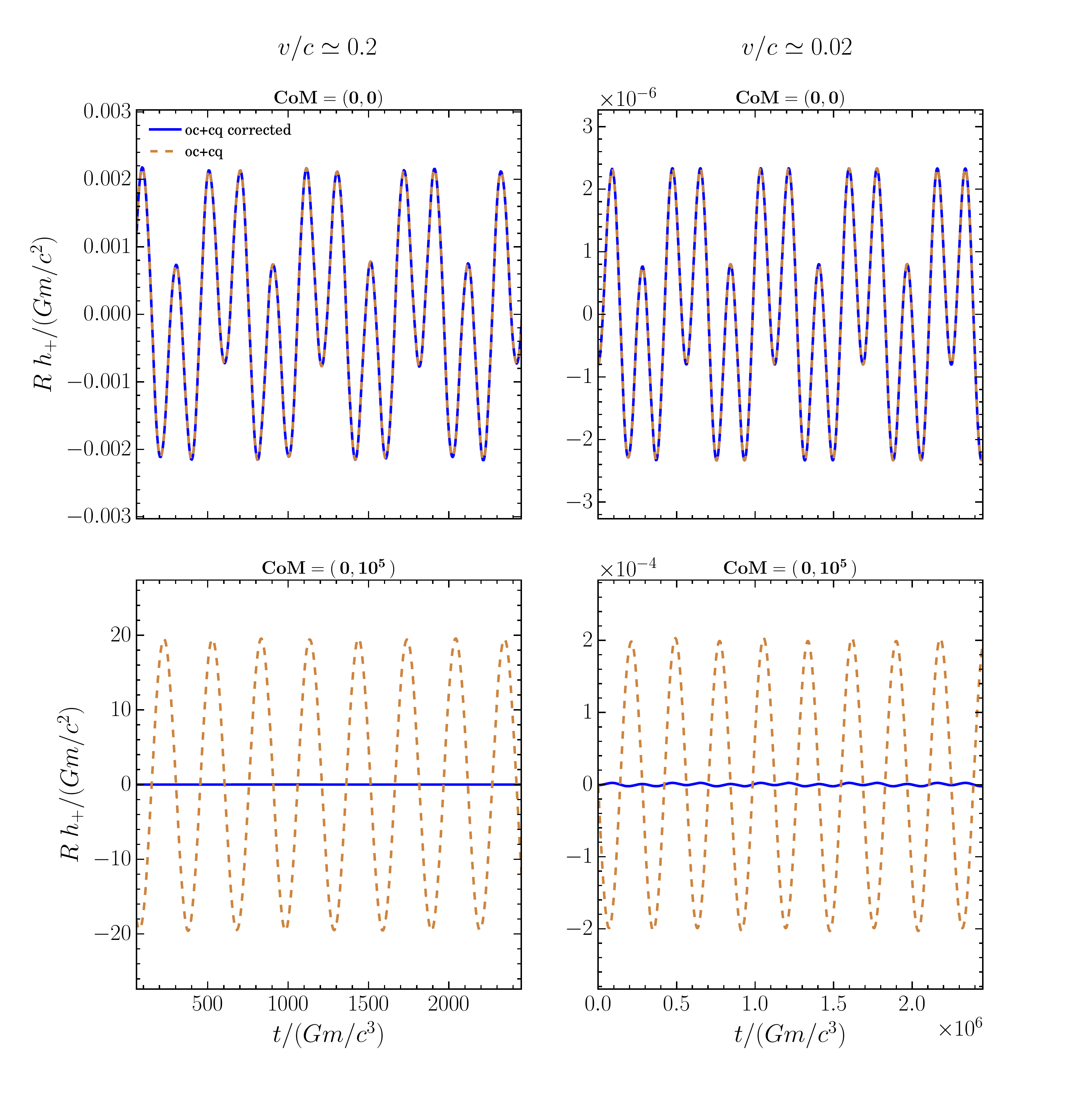}
		\caption{Same as in fig.~\ref{fig:quad_corrected}, except that we
          report the waveforms computed with the mass octupole and current quadrupole
          corrections (at the 0.5PN order) whereas the observer is located on the
          $y$ axis. Again, note the different $y$-axis scales between the
          upper and lower panels for each value of $v/c$.}
	\label{fig:oc_cq_corrected}
	\end{figure}
\end{center}

Let us consider now the next-to-leading order contributions to the waveform,
comprised of a mass octupole and a current quadrupole term. When these terms
are taken into account, eq.~\ref{eq:first_quad_formula} becomes (see~\cite{Maggiore2007_book} \S 3.4)
%
\begin{equation}\label{eq:h_oct}
\fl h_{ij}^{\rm TT}(t,\bm{x}) = \frac{2 G}{R c^4}\Lambda_{ijkl}(\bm{n}) \biggl[
  \frac{\ud^2 M^{kl}}{\ud t^2}
+ \frac{n_m}{3c}\Big( \frac{\ud^3 M^{klm}}{\ud t^3} + 2 \frac{\ud^2 S^{klm}}{\ud
  t^2}\Big) \biggr] + \mathcal{O}\Big(\frac{1}{c^6}\Big)\, ,
\end{equation}
where $\bm{n} = \bm{x}/R$, and $M^{klm}, S^{klm}$ are respectively the
Newtonian octupole and current quadrupole moments (evaluated at $t_{\rm ret}$
[see \ref{sec:app_A}]). Again, for a binary in the $xy$ plane and an observer in
the same plane\footnote{We choose the observer in the $xy$ plane rather than
  in the $z$ direction because the mass octupole and current quadrupole
  corrections vanish with the latter choice.}, the above equation implies
(see Problem 3.3 of~\cite{Maggiore2007_book} and the corresponding erratum)
%
\begin{eqnarray}\label{eq:h_oct_bin}
(h_+)_{\rm oc+cq} 
&= \frac{1}{R} \frac{G^{5/2} m_1 m_2 \delta m}{4 (m
  a^3)^{1/2} c^5}  \left[5\cos(\omega t_{\rm ret}) -
  9\cos(3\omega t_{\rm ret}) \right], \nonumber\\
  (h_{\times})_{\rm oc+cq} &= 0\, ,
\end{eqnarray}
with $\delta m = m_1-m_2$. For the
two binaries considered above, we obtain
$Rc^2/(Gm) (h_+)_{\rm oc+cq}\simeq 2.2\times 10^{-3}$ in the relativistic
case, and $Rc^2/(Gm) (h_+)_{\rm oc+cq} \simeq 2.2\times 10^{-6}$ in the
mildly relativistic one.
From fig.~\ref{fig:oc_cq_corrected} (dashed lines), we can see that if the
origin of the coordinates coincides with the CoM, then $(h_+)_{\rm oc+cq}$ as
given by eq.~\ref{eq:h_oct} applied to the numerically-integrated orbits
agrees well with the analytic result of eqs.~\ref{eq:h_oct_bin}. Conversely, if
one shifts significantly the origin of the coordinates, $(h_+)_{\rm oc+cq}$
computed from the numerical trajectories no longer agrees with
eqs.~\ref{eq:h_oct_bin}, even for the mildly-relativistic binary.

In the next two sections, we will analyze the reasons behind these
discrepancies and explain how they can be avoided, first for the Newtonian
quadrupole formula (sec.~\ref{sub:quad}), next for the 0.5PN quadrupole formula
with octupolar corrections (sec.~\ref{sub:oct}).

\subsection{Quadrupole waveform} 
\label{sub:quad}

If we explicitly compute the second-order time derivatives in
eq.~\ref{eq:first_quad_formula}, for a binary system, we obtain
%
\begin{eqnarray}\label{eq:h_der_exp}
h_{ij}^{\rm TT}(t,\bm{x}) 
& = \frac{2 G}{R c^4} \Lambda_{ijkl}(\bm{n})\biggl( 2 m_1
  \dot{x}_1^k \dot{x}_1^l + 2 m_2 \dot{x}_2^k \dot{x}_2^l \nonumber\\
&+ m_1 (\ddot{x}_1^k x_1^l + x_1^k \ddot{x}_1^l) + m_2 (\ddot{x}_2^k x_2^l + x_2^k \ddot{x}_2^l) \biggr) +
  \mathcal{O}\Big(\frac{1}{c^5}\Big)\, .
\end{eqnarray}
Now, the position vectors of the two masses can be expressed in terms of the
CoM position $\bm{r}_{0}$ and the relative separation vector
$\bm{r}=\bm{x}_1-\bm{x}_2$ as
%
\begin{equation}\label{eq:cm_def}
\bm{x}_1 = \bm{r}_{0} + \frac{m_2}{m}\bm{r}\, , \qquad \qquad 
\bm{x}_2 = \bm{r}_{0} - \frac{m_1}{m}\bm{r}\, ,
\end{equation}
so that eq.~\ref{eq:h_der_exp} takes the form
%
\begin{eqnarray}\label{eq:h_cm_sub}
 \fl  h_{ij}^{\rm TT}(t,\bm{x}) 
  &= \frac{2 G}{R c^4} \Lambda_{ijkl}(\bm{n}) \bigg[ \frac{2}{m} (m_1 \dot{x}_1^k+m_2
    \dot{x}_2^k)(m_1 \dot{x}_1^l+m_2 \dot{x}_2^l) \nonumber \\ \fl & + \frac{2m_1 m_2}{m} 
(\dot{x}_1^k-\dot{x}_2^k)(\dot{x}_1^l-\dot{x}_2^l) + 
r_0^{k}(m_1 \ddot{x}_1^{l} + m_2 \ddot{x}_2^{l}) + 
r_0^{l}(m_1 \ddot{x}_1^{k} + m_2 \ddot{x}_2^{k})\nonumber\\  
\fl   & + \frac{m_1 m_2}{m} \Big( r^k(\ddot{x}_1^l-\ddot{x}_2^l) + r^l(\ddot{x}_1^k-\ddot{x}_2^k)\Big)
    \bigg] + \mathcal{O}\Big(\frac{1}{c^5}\Big)\, .
\end{eqnarray}
Since the CoM absolute coordinates $\bm r_0$ explicitly appears in this
expression, it would seem that the GW amplitude should depend on the choice of
the origin of the coordinate system. However, because eq.~\ref{eq:h_cm_sub} is only
correct at leading order in PN theory, it is actually sufficient to compute the
accelerations $\bm{\ddot{x}}_1$ and $\bm{\ddot{x}}_2$ at leading (i.e., Newtonian) order.
If one does so, the identity $m_1 \bm{\ddot{x}}_1 + m_2 \bm{\ddot{x}}_2=0$ holds for an
isolated system, hence the dependence on the position of the center of mass
(and thus on the location of the origin) disappears from
eq.~\ref{eq:h_cm_sub}. Similarly, $m_1 \dot{x}_1^k+m_2 \dot{x}_2^k$ is constant and
independent of the location of the origin.

One may want, however, to integrate the binary's equations of motion to higher
PN order, either analytically or numerically. For instance, the code of
Ref.~\cite{Bonetti2016} integrates the PN Hamiltonian for binary or triple
systems through the 2.5PN order including the dissipative effects of radiation
reaction. Now, when one includes these PN corrections,
$m_1 \bm{\ddot{x}}_1 + m_2 \bm{\ddot{x}}_2\neq0$ already at 1PN order, so that the
dependence on the location of the origin does \textit{not} disappear. This is
the reason of the unphysical behavior visible in fig.~\ref{fig:quad_corrected}
(and partly in fig.~\ref{fig:oct_triple0}, see also the next section).

A first solution can be therefore to avoid the use of the numerical
trajectories to compute the accelerations in eq.~\ref{eq:h_cm_sub}, but
instead evaluate them directly from the positions of the two bodies by using the
Newtonian dynamics (i.e.,~Newton's second law, which ensures
$m_1 \bm{\ddot{x}}_1 + m_2 \bm{\ddot{x}}_2=0$). Alternatively, one can note that the
combination $m_1 \bm{\ddot{x}}_1 + m_2 \bm{\ddot{x}}_2$ is simply (at Newtonian order) the
time derivative of the total linear momentum
$\bm{P}_{\rm N}=m_1 \bm{\dot{x}}_1 + m_2 \bm{\dot{x}}_2$. Thus, the identity
$m_1 \bm{\ddot{x}}_1 + m_2 \bm{\ddot{x}}_2=0$ just reflects the conservation of
$\bm{P}_{\rm N}$, which is an automatic consequence of the Newtonian dynamics.
Beyond it, when PN corrections are included, $m_1 \bm{\ddot{x}}_1 + m_2 \bm{\ddot{x}}_2$ does
not vanish, because the Newtonian linear momentum
$\bm{P}_{\rm N}=m_1 \bm{\dot{x}}_1 + m_2 \bm{\dot{x}}_2$ is no longer a conserved
quantity. This is what causes the dependence on the choice of $\bm{r}_0$
observed in fig.~\ref{fig:quad_corrected}. However, one can exploit the fact
that there exists a conserved PN linear momentum $\bm{P}_{n {\rm PN}}$
generalizing $\bm{P}_{\rm N}$ at the $n$PN order. In practice, replacing
$\bm{P}_{\rm N}$ with $\bm{P}_{n {\rm PN}}$ is equivalent to computing the
accelerations appearing in eq.~\ref{eq:h_cm_sub} as $\bm{\ddot{x}}_i=\bm{\pi}_i/m_i$,
with $i=1,2$ and $\bm{\pi}_i$ denoting the conjugate momentum of each body
entering the Hamilton equations. Then, the combination
$m_1 \bm{\ddot{x}}_1 + m_2 \bm{\ddot{x}}_2$ always vanishes, even if PN corrections are
included in the Hamiltonian dynamics.

In conclusion, either of these two workarounds (which give rise to the
``amended'' waveforms represented by blue solid lines in
fig.~\ref{fig:quad_corrected}) is sufficient to eliminate the unphysical
dependence on the origin of the coordinates.

\subsection{Octupole and Current Quadrupole waveforms}
\label{sub:oct}

Let us now examine what happens to the contribution of the mass octupole and
current quadrupole moments to the waveform under a change of reference frame
of the form of eq.~\ref{eq:cm_def}. After expanding the time derivatives
appearing in the term $n_m/(3c)\, (\tdot{M}^{klm} \!\!+ 2 \ddot{S}^{klm})$ of
eq.~\ref{eq:h_oct} by means of the Leibniz rule, the dependence on the CoM
location does not cancel out in the waveform, but gives instead a
contribution\footnote{Note that we have neglected terms $\propto \delta^{ij}$
  in the sum as they disappear when they are TT-projected since
  $\Lambda_{ijkl}(\bm{n})\delta^{kl}=0$ (see~\ref{sec:app_A}).}
%
\begin{center}
	\begin{figure}
		\centering
		\includegraphics[scale=0.45]{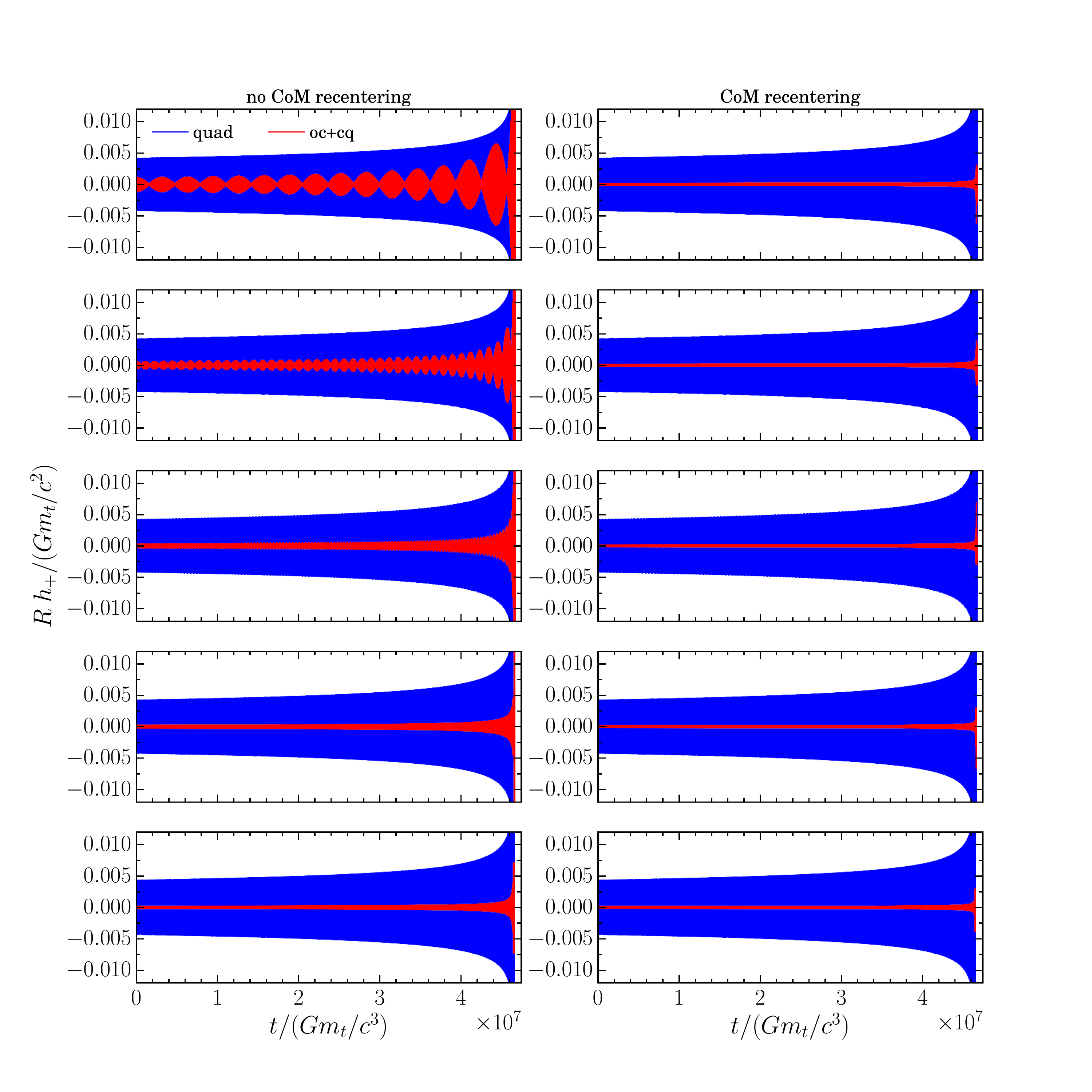}
		\caption{Same triplets as in fig.~\ref{fig:oct_triple0}. {\it Left}:
          Waveform computed in the frame of the triplet's CoM. {\it Right}:
          Waveform computed after shifting the origin to the CoM of
          the inner binary. Again to be compared to~\cite{Galaviz2011},
          fig.~18.}
		\label{fig:oct_triple}
		\end{figure}
\end{center}

\begin{equation}\label{eq:offending_terms}
\fl  \delta_{r_0} h_{ij}^{\rm TT} = \Lambda_{ijkl}(\bm{n})\frac{n_m r_0^m}{3c}\sum_A m_A
  \left(3 \tdot{x}_A^k x_A^l + 3 
    x_A^k \tdot{x}_A^l + 9 \ddot{x}_A^k \dot{x}_A^l + 9 \dot{x}_A^k
    \ddot{x}_A^l\right)\, .
\end{equation}
This may be rewritten as
%
\begin{equation}\label{eq:d3Q}
\delta_{r_0} h_{ij}^{\rm TT}=\Lambda_{ijkl}(\bm{n})\frac{\ud^3 M^{kl}}{\ud
  t^3}\frac{\delta R}{c}\, , 
\end{equation}
with $\delta R = \bm{n} \cdot \bm{r}_0 = n_m r_0^m$. Hence, the terms
proportional to $\bm{r}_0$ produced by the mass octupole and current
quadrupole moments can be reabsorbed in a shift $\delta R/c$ of the retarded
time at which the (quadrupole) waveform is evaluated:

\begin{eqnarray} \label{eq:resum}
\fl  \Lambda_{ijkl}(\bm{n}) \bigg[ \frac{\ud^2 M^{kl}(t-R/c)}{\ud t^2}  &+
\frac{\ud^3 M^{kl}(t-R/c)}{\ud t^3} \frac{\delta R}{c}\bigg] \nonumber
\\ \fl &= \Lambda_{ijkl}(\bm{n}) \bigg[\frac{\ud^2
         M^{kl}(t-R/c+\delta R/c)}{\ud t^2}  \bigg] +
         \mathcal{O}\bigg(\frac{[\delta R]^2}{c}\bigg) \, . 
\end{eqnarray}

This time shift simply enforces the invariance of the waveform under
translations of the reference frame in which it is computed. Indeed, the
retarded time is always, by definition, $t_{\rm ret} = t-|\mathbf{x}|/c$ in
the generic frame $(t,\bm{x})$ (assuming radiative coordinates), where
$R=|\bm{x}|$ is the distance between the observer at position $\bm{x}$ and the
origin. This implies in particular that
$t_{\rm ret}-t_{\rm ret}^{\rm CoM} = |\bm{x}^{\rm CoM}|/c-|\bm{x}|/c = -\delta
R/c + \mathcal{O}([\delta R]^2/c)$,
where the quantities labeled with the superscript CoM are referring to the CoM
frame. On the other hand, the expressions of the multipole moments when
$\bm{r}_0\neq \bm{0}$ differ from their standard forms in CoM coordinates.
Obviously, the modifications of $t_{\rm ret}$ and those of the multipole
moments must (and do!) compensate each other so that the GW signal remains
invariant, irrespective of the choice of the origin of the coordinates. This
main conclusion remains true for linearly propagating waves when other
multipole moments are taken into account\footnote{Beyond linear order, one
  must replace the source moments $M_{ab}$, $M_{abc}$, ..., $S_{abc}$ in
  eq.~\ref{eq:hinvariance} by the so-called ``radiative moments'' which
  parametrize the gravitational waveform.}:
%
\begin{equation} \label{eq:hinvariance} \fl
h_{ij}^{\rm TT} [\{M_{ab},M_{abc},...,S_{abc},...\},t,\bm{x}] = h_{ij}^{\rm TT}
[\{M_{ab}^{\rm CoM},M_{abc}^{\rm CoM},...,S_{abc}^{\rm
  CoM},...\},t,\bm{x}^{\rm CoM}]\, .
\end{equation}

However, the analytic ``resummations'' needed for the above argument to work,
such as the one in eq.~\ref{eq:resum}, are based on a Taylor expansion.
Therefore, one has to implicitly assume that terms like those in
eq.~\ref{eq:offending_terms} are ``small'' or, more precisely, that the
displacement $| \bm{r}_0|$ of the CoM from the origin of the coordinates is
much smaller than the wavelength of the quadrupole waveform,
$\lambda=\pi c/\omega$. This is in fact very natural because, as already
mentioned, one of the assumptions implicit in the derivation of the
quadrupole/octupole formulae is that the source be well contained in a NCZ of
size $\sim \lambda$ centered on the origin of the coordinates, where
retardation effects are negligible. If the generalized quadrupole
formula is applied to systems for which the source is \textit{not} well
contained in its NCZ, one will \textit{not} be able to resum the terms of
eq.~\ref{eq:offending_terms} into a time shift. This is the origin of the
discrepancy shown in fig.~\ref{fig:oc_cq_corrected} for binary systems.

\begin{figure}
 \begin{minipage}[c]{0.43\textwidth}
   \centering
   \includegraphics[scale=0.2]{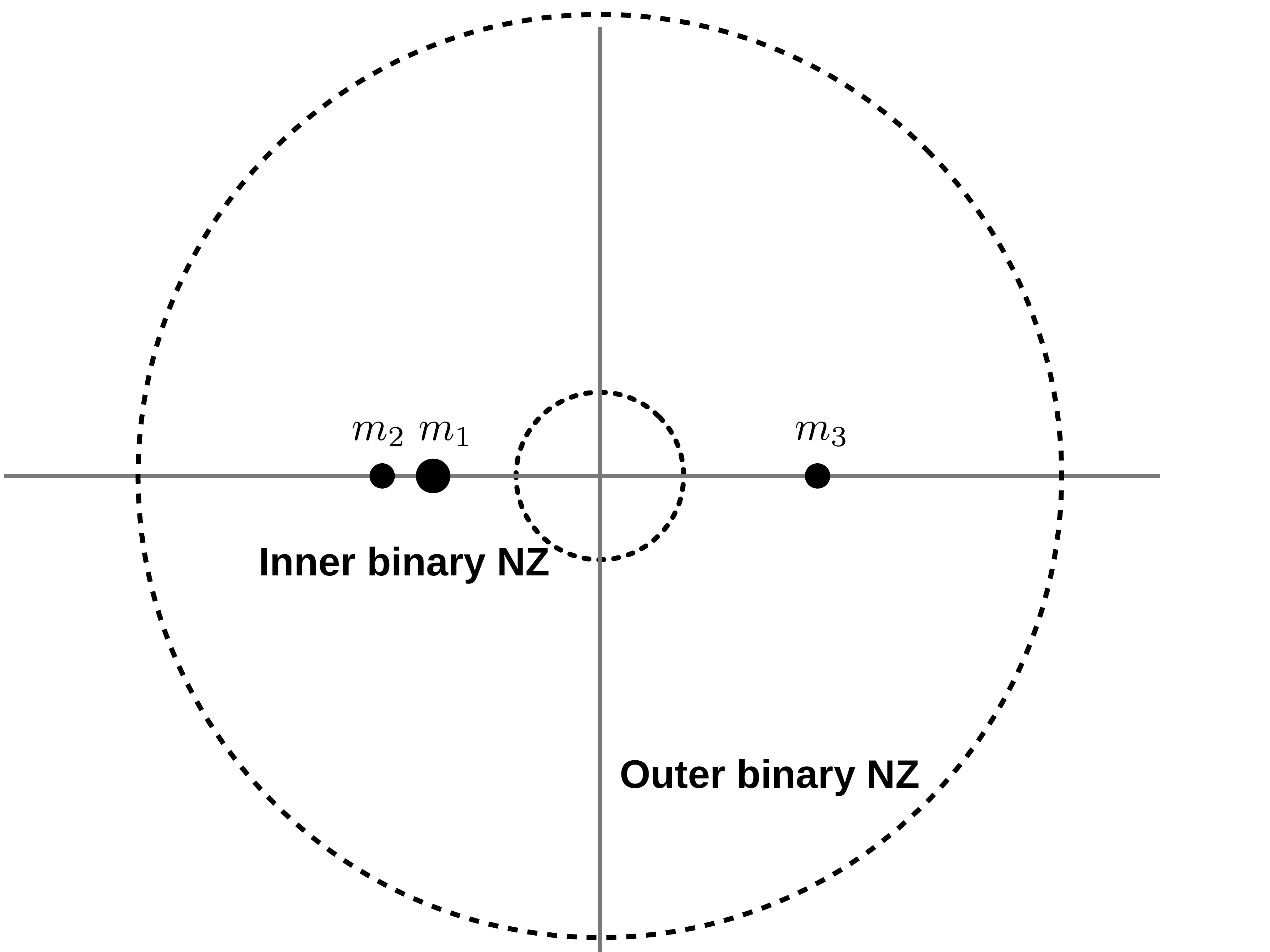}
 \end{minipage}
 \ \hspace{2mm} \hspace{3mm} \
 \begin{minipage}[c]{0.43\textwidth}
  \centering
   \includegraphics[scale=0.2]{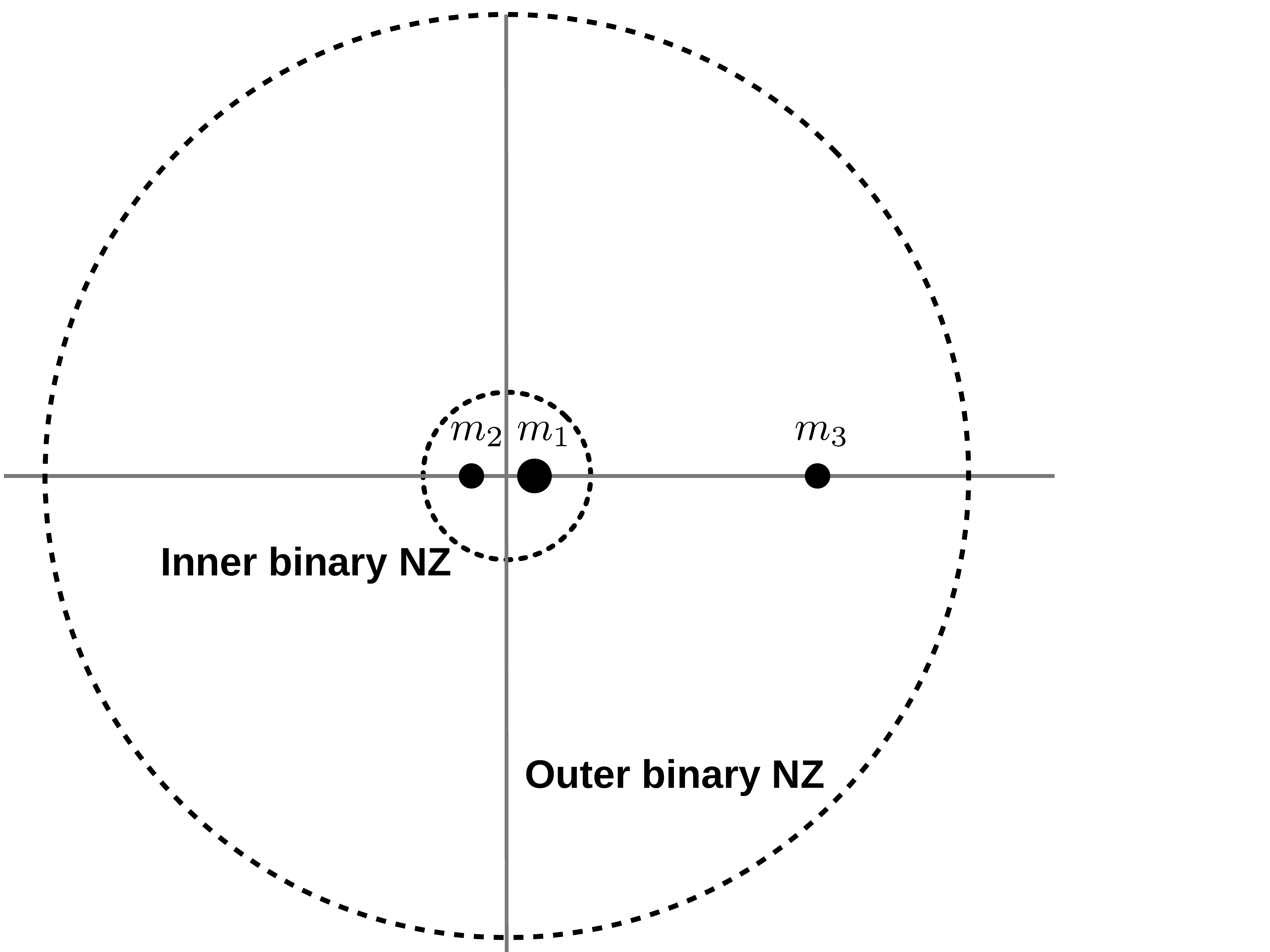}
    \end{minipage}
    \caption{Cartoon representation of the change of reference frame needed to
      fix the unphysical spurious behavior shown in
      fig.~\ref{fig:oct_triple0}. {\it Left panel}: the origin of the
      reference frame coincides with the CoM of the triple system. {\it Right
        panel}: the origin instantaneously coincides with the inner binary's
      CoM. The latter choice allows both the inner and outer binaries to lie
      well within their respective NCZs.}
  \label{fig:NZcartoon}
\end{figure}

This observation also highlights the reason of the unphysical behavior shown
in fig.~\ref{fig:oct_triple} (left panels) for \textit{triple} systems. In
fact, for a weakly/mildly relativistic binary ($v\lesssim c$) with separation
$a$, one has $\lambda \sim a /(v/c) \gtrsim a$, i.e., if one chooses the
origin of the coordinates to coincide with the binary's CoM, the NCZ will
always contain the binary. For a hierarchical triple system, instead, if one
sets the origin at the location of the triplet's CoM, the inner binary will be
\textit{outside} its NCZ provided that the separation of the outer binary is
sufficiently large. This may be understood by noting that there are actually two
NCZs for a hierarchical triplet, i.e., an inner-binary NCZ with size
$\lambda_{\rm in}\sim a_{\rm in}/(v_{\rm in}/c)$, and an outer-binary NCZ with
size $\lambda_{\rm out}\sim a_{\rm out}/(v_{\rm out}/c)$, as illustrated in
fig.~\ref{fig:NZcartoon}. Clearly, while the outer binary will always be
contained in its NCZ if $v_{\rm out}/c\lesssim 1$, the inner binary will
eventually be outside its NCZ if $a_{\rm out}$ is sufficiently large. Indeed,
as $a_{\rm out}$ increases, the inner binary's CoM ends up leaving its own NCZ
(which is centered on the origin of the coordinates, i.e. on the triplet's
CoM).

A simple fix to this issue, as shown in fig.~\ref{fig:NZcartoon}, is thus to
evaluate the multipole moments in an \textit{inertial} reference frame with
origin instantaneously coinciding with the CoM of the \textit{inner
  binary}\footnote{Clearly, this reference frame cannot be co-moving with the
  CoM of the inner binary (which has a non-zero acceleration) and one has
  to consider a different inertial frame at each step of the system's
  evolution.}, which allows both the inner and outer binaries to lie within
their respective NCZs. In the right panels of fig.~\ref{fig:oct_triple}, we
show that this eliminates the unphysical behavior of
fig.~\ref{fig:oct_triple0} and left panels of fig.~\ref{fig:oct_triple}.
Therefore, the problem exhibited by those figures (and the corresponding
results of fig.~18 in~\cite{Galaviz2011}) was simply that the waveforms were
evaluated in the reference frame of the CoM of the triple system. The very
same solution applies to the simpler binary cases reported in
fig.~\ref{fig:oc_cq_corrected}. Here, since the CoM does not move, only a
single transformation is needed. Once that transformation has been performed,
the waveforms are given by the solid blue lines and reproduce the correct
predicted result.

\section{Green's solution in linearized theory}
\label{sec:green_solution}

\setcounter{footnote}{0}

As pointed out in the introduction, another often overlooked problem arises when
gravitational waveforms are computed by direct integration of
eq.~\ref{eq:first_wave_eq} with the help of the retarded Green function. To understand
it, let us go back to the textbook derivation of the quadrupole
formula for GW generation. We start from the Einstein
equations relaxed by the condition of harmonic coordinates {(see,
  e.g.,~\cite{MTW} \S 20.3, \S 36.9, or \cite{Blanchet2014} \S 6.3
  and~\cite{Pati2000} for the complete derivation):
%
\begin{equation}
\label{eq:relaxed_eq}
\Box_{\rm flat} H^{\alpha\beta}  = -\frac{16 \pi G}{c^4}\tau^{\alpha\beta}\, ,
\end{equation}
where the pseudo-tensor $H^{\alpha\beta}$ is defined in terms of the Minkowski
metric $\eta^{\alpha\beta}$ and the space-time (inverse) metric
$g^{\alpha\beta}$ as
%
\begin{equation}\label{eq:def_H}
H^{\alpha\beta} \equiv \eta^{\alpha\beta} - (-g)^{1/2}g^{\alpha\beta}\,,
\end{equation}
and satisfies the harmonic gauge condition, i.e.,
$\partial_{\beta} H^{\alpha\beta} = 0$. Moreover, the ``effective''
stress-energy pseudo-tensor $\tau^{\alpha\beta}$ is comprised of a
contribution from the stress-energy tensor of matter, and a contribution
$\Lambda^{\alpha\beta}$ from the non-linearities of the gravitational field,
i.e.,
%
\begin{equation}\label{eq:def_tau}
\tau^{\alpha\beta} = (-g)T^{\alpha\beta} + \frac{c^4}{16\pi G}
\Lambda^{\alpha\beta}\, ,
\end{equation}
where
%
\begin{equation}
\Lambda^{\alpha \beta}
 = \frac{16\pi G}{c^4} (-g) t_{\rm LL}^{\alpha \beta }  +
 ( \partial_\nu  H^{\alpha \mu} \partial_\mu H^{\beta \nu}
 -\partial_{\mu}\partial_{\nu} H^{\alpha \beta} H^{\mu \nu} ) \,,
\end{equation}
with $ t_{\rm LL}^{\alpha \beta }$ denoting here the Landau-Lifshitz
pseudo-tensor~\cite{Landau_book1975}. As a consequence of the harmonic gauge
condition, $\tau^{\alpha\beta}$ is also flat-space conserved, i.e.,
$\partial_{\beta} \tau^{\alpha\beta} = 0$.

By using the retarded Green function, we can now integrate
eq.~\ref{eq:relaxed_eq} and obtain its formal solution:
%
\begin{equation}\label{eq:H_local_ret}
H^{\alpha\beta}(t,\bm{x}) = \frac{4G}{c^4}\int
\frac{\ud^3\bm{x}'}{|\bm{x}-\bm{x}'|}\tau^{\alpha\beta}
\left(t-|\bm{x}-\bm{x}'|/c,\bm{x}'\right)\,.
\end{equation} 
If the field point $\bm{x}$ lies very far from the source,
$|\bm{x}-\bm{x}'|\approx |\bm{x}|\equiv R$, and one can neglect the
differences in retarded time among the source components by considering a
single global retardation $t_{\rm ret}=t-R/c$. This yields\footnote{When
  neglecting the $\bm{x}'$ term in the temporal dependency of the
  integral in eq.~\ref{eq:H_local_ret} and replacing the source $\tau^{\alpha\beta}$
  by its PN expansion, the integral on the right-hand side of
  eq.~\ref{eq:H_global_ret} becomes formally divergent. This is one of the
  main problems that  GW generation formalisms have to address. In the
  Blanchet-Damour-Iyer formalism~\cite{BDI1995,BDIWW1995}, this particular problem is solved by
  resorting to a combination of asymptotic matching techniques and a specific
  regularization procedure that cures those divergences
  \cite{BD86,BD89,B98mult,PB02}.}
%
\begin{equation}\label{eq:H_global_ret}
H^{\alpha\beta}(t,\bm{x}) \approx \frac{4G}{R c^4}\int
\ud^3\bm{x}'\tau^{\alpha\beta}\left(t_{\rm ret},\bm{x}'\right)\, .
\end{equation}
Let us now expand the metric up to $1/c^2$ corrections,
%
\begin{eqnarray}\label{eq:phi}
  g_{00} &= -1 -2\frac{\phi}{c^2} + \mathcal{O}\Big(\frac{1}{c^4}\Big),\nonumber\\
  g_{0i} &= \mathcal{O}\Big(\frac{1}{c^3}\Big),\nonumber\\
  g_{ij} &= \Big(1-2\frac{\phi}{c^2}\Big)\delta_{ij} 
+ \mathcal{O}\Big(\frac{1}{c^4}\Big)\, ,
\end{eqnarray}
so that the $ij$ components are given at the 1PN order while the $00$ and $0i$
components are Newtonian. The gravitational potential $\phi$ in
eqs.~\ref{eq:phi} must satisfy the Poisson equation
$\nabla^2 \phi = 4\pi G \rho$, with $\rho$ being the mass density, in order
for $\phi$ to be a solution of the relaxed Einstein equations. We can see that,
at this accuracy level, the metric is linear in the source, which implies
%
\begin{equation}
H^{\mu\nu} = \bar{h}^{\mu\nu} +
\mathcal{O}\Big(\frac{1}{c^4},\frac{1}{c^3},\frac{1}{c^4}\Big)\, ,
\end{equation}
where the three remainders in the arguments of the Landau symbol refer to the
$00$, $0i$ and $ij$ components, respectively. From the
flat-space conservation of $\tau^{\mu\nu}$, it then follows
that (see~\cite{MTW} \S 36.10)\footnote{The very same identity holds for $T^{\mu\nu}$ in
  linearized theory, since $\partial_{\mu}T^{\mu\nu}=0.$}
%
\begin{eqnarray}\label{eq:MTW_identity}
\partial_0\partial_0(\tau^{00} x^j x^k) =\partial_l\partial_m(\tau^{lm} x^j
  x^k) - 2\partial_l(\tau^{jl} x^k + \tau^{kl} x^j) + 2 \tau^{jk}\, ,
\end{eqnarray}
which allows one to recast the spatial part of eq.~\ref{eq:H_global_ret} as (see~\cite{MTW} \S 36.10)
%
\begin{equation}\label{eq:hij}
\bar{h}^{ij}(t,\bm{x}) \approx \frac{2G}{R c^4} \frac{\ud^2}{\ud t^2}\int
\ud^3\bm{x}' c^{-2} \tau^{00}\left(t_{\rm ret},\bm{x}'\right)x'^ix'^j  +
\mathcal{O}\Big(\frac{1}{c^5}\Big)\,.
\end{equation}
A priori, the PN expansion of the stress-energy pseudo-tensor $\tau^{\alpha\beta}$
defined by eq.~\ref{eq:def_tau} still involves both the matter term $T^{\alpha\beta}$ and
the purely gravitational non-linear source term $\Lambda^{\alpha\beta}$.
Modeling the matter system by point particles, i.e., taking (see~\cite{Weinberg1972book} \S 2.8,\cite{Maggiore2007_book} \S 3.3.5)
%
\begin{equation}\label{eq:T_point}
T^{\alpha\beta}(t,\bm{x}) = \sum_A 
\frac{m_A u_A^{\alpha} u_A^{\beta}}{(u^0_A/c) \
  \sqrt{-g}}\delta^3(\bm{x}-\bm{x}_A(t))\, ,
\end{equation}
where $m_A$ is the mass of particle $A$ and $u_A^\alpha$ is its four-velocity,
one finds
%
\begin{eqnarray}\label{eq:tau_expansion_a}
\tau^{00}(t,\bm{x}) 
&= \sum_A m_A c^2 \delta^3\left(\bm{x}-\bm{x}_A(t)\right) +
  \mathcal{O}\Big(\frac{1}{c^0}\Big)  \, ,\\ \label{eq:tau_expansion_b}
  \tau^{0i}(t,\bm{x})
&= \sum_A m_A c \, \dot{x}_A^i \delta^3\left(\bm{x}-\bm{x}_A(t)\right) +
  \mathcal{O}\Big(\frac{1}{c}\Big) \, ,\\
  \label{eq:tau_expansion_c}
  \tau^{ij}(t,\bm{x})
&= \sum_A m_A \dot{x}_A^i \dot{x}_A^j \delta^3\left(\bm{x}-\bm{x}_A(t)\right)
   \nonumber\\
& + \frac{1}{4\pi
  G}\Big(\partial^i\phi\, \partial^j\phi-
  \frac{1}{2}\delta^{ij}\partial_k\phi \, \partial^k\phi\Big)
  + \mathcal{O}\Big(\frac{1}{c^2}\Big)\, ,  
\end{eqnarray}
where $\delta^3$ is the three-dimensional Dirac delta function and $\dot{x}_A^i$
represents the components of the three-dimensional velocity. We observe that
$\tau^{00}\approx T^{00}$ at leading PN order, hence eq.~\ref{eq:hij}
coincides with the ``usual'' quadrupole formula,
%
\begin{equation}\label{eq:4polo}
\bar{h}^{ij}(t,\bm{x}) \approx \frac{2 G}{R c^4} \frac{\ud^2 M^{ij}}{\ud
  t^2}(t_{\rm ret}) + \mathcal{O}\Big(\frac{1}{c^5}\Big)\, .
\end{equation}
%
By contrast, $\tau^{ij}$ contains a direct contribution from the gravitational
field at Newtonian order, i.e., a term arising at the same order as
$T^{ij}$. Therefore, the direct integration of
eq.~\ref{eq:first_wave_eq} will give an \textit{incorrect} result, because the
right-hand side is wrong already at the leading PN order.

To be more explicit, let us evaluate eq.~\ref{eq:hij} in the special
case of a binary system ($A,B=1,2$) by using eq.~\ref{eq:tau_expansion_a} and
the Newtonian equations of motion
\begin{equation}
\ddot{x}_A^i = -\sum_{B \neq A} \frac{G m_B }{r_{AB}^2}n^i_{AB}\, ,
\end{equation}
where $\bm{r}_{AB} = \bm{x}_A - \bm{x}_B$, $r_{AB} = |\bm{x}_A - \bm{x}_B|$, and
$\bm{n}_{AB} = \bm{r}_{AB}/r_{AB}$. Straightforward algebra yields
\begin{equation}\label{eq:h_2b_quadrupole}
\bar{h}^{ij}(t,\bm{x}) \approx \frac{4 G}{R c^4}\Big(m_1 \dot{x}_1^i \dot{x}_1^j + 
m_2 \dot{x}_2^i \dot{x}_2^j -\frac{G m_1 m_2 n_{12}^i n_{12}^j}{r_{12}}\Big) + 
\mathcal{O}\Big(\frac{1}{c^5}\Big)\, ,
\end{equation}
while direct integration of eq.~\ref{eq:first_wave_eq} would lead to the
different expression
\begin{equation}\label{eq:h_green}
\bar{h}^{ij}(t,\bm{x}) \approx \frac{4 G}{R c^4}\Big( m_1 \dot{x}_1^i \dot{x}_1^j + 
m_2 \dot{x}_2^i \dot{x}_2^j \Big) + \mathcal{O}\Big(\frac{1}{c^5}\Big)\, .
\end{equation}

The extra term in eq.~\ref{eq:h_2b_quadrupole} that is missing in
eq.~\ref{eq:h_green} is related to the purely gravitational part of the
right-hand side of eq.~\ref{eq:tau_expansion_c} (i.e., the part involving the
Newtonian potential $\phi$), which is prematurely neglected in the source
$T^{ij}$ of eq.~\ref{eq:first_wave_eq}. This is the origin of the factor
$\sim 2$ discrepancy shown in fig.~\ref{fig:quad_vs_green}.

Nonetheless, one can still obtain the correct expression without resorting to
the identity~\ref{eq:MTW_identity}. In fact, by substituting
eq.~\ref{eq:tau_expansion_c} into the spatial components of
eq.~\ref{eq:H_global_ret}, one gets
%
\begin{eqnarray}
\label{eq:H_spatial}
\fl \bar{h}^{ij}(t,\bm{x}) \approx \frac{4G}{c^4} 
& \int \frac{\ud^3\bm{x}'}{|\bm{x}-\bm{x}'|}\biggl[ T^{ij} + 
\frac{1}{4\pi G} \Big(\partial^i\phi\, \partial^j\phi-
\frac{1}{2}\delta^{ij}\partial_k\phi\, \partial^k\phi\Big)\biggr] + 
\mathcal{O}\Big(\frac{1}{c^5}\Big)\,,
\end{eqnarray}
which may be rewritten as (see \ref{sec:app_C} for details)

\begin{eqnarray}\label{eq:h_cont}
\fl \bar{h}^{ij}(t,\bm{x}) \approx
& \frac{4G}{R c^4}\int \ud^3\bm{x}' \ T^{ij}- 
\frac{2 G^2}{R c^4}\int \ud^3\bm{y}' \ud^3\bm{y}''\rho(\bm{y}')\rho(\bm{y}'') 
\frac{\hat{n}^i \hat{n}^j}{|\bm{y}'-\bm{y}''|} +
  \mathcal{O}\Big(\frac{1}{c^5}\Big)\, , 
\end{eqnarray}
where $\hat{n}^{k} = (y'^{k}-y''^{k})/|\bm{y}'-\bm{y}''|$. 
Focusing again on the case of two point particles, eq.~\ref{eq:h_cont} gives
%
\begin{equation}
\fl \bar{h}^{ij}(t,\bm{x}) \approx \frac{4G}{R c^4}
\Big(m_1 \dot{x}_1^i \dot{x}_1^j + m_2 \dot{x}_2^i \dot{x}_2^j - 
\frac{ G m_1 m_2 n_{12}^i n_{12}^j}{r_{12}}\Big) + \mathcal{O}\Big(\frac{1}{c^5}\Big)\,.
\end{equation}
This expression agrees with eq.~\ref{eq:h_2b_quadrupole}. In conclusion,
  the discrepancy shown in fig.~\ref{fig:quad_vs_green} was simply due to
  neglecting the purely gravitational part of the source in the relaxed
  Einstein equations, which would be equivalent to assuming motion along
  straight lines for the binary components. In other words, the last term in
  eq. \ref{eq:h_2b_quadrupole} accounts for the fact that motion does not take
  place on rectilinear trajectories, but instead along curved-spacetime
  geodesics.
  
\section{Conclusions}
\label{sec:conclusions}

In this paper, we have highlighted several subtleties that occur when applying
the quadrupole and quadrupole-octupole formulae to numerically-integrated
binary and triple systems. We have shown that, as expected, applying these
formulae to a binary in a reference frame whose origin is displaced from the
CoM of the system leads to unphysical spurious results, if the displacement
exceeds the wavelength $\lambda$ of the emitted GWs. This simply happens
because implicit in the derivation of the (generalized) quadrupole formula is
the assumption that the multipoles are defined in the NCZ~\cite{Blanchet2014},
i.e., a region of size $\sim\lambda$ centered on the origin in which the
binary is supposed to be contained. The same problem manifests itself in
hierarchical triple systems, when the quadrupole and quadrupole-octupole
formulae are applied in a reference frame centered on the triplet's CoM. The
resulting unphysical behavior, which to the best of our knowledge has gone
unrecognized in the literature~\cite{Galaviz2011}, is in contrast with what
happens for a binary system (where it is safe to define the multipoles in the CoM
reference frame), but can be understood by bearing in mind that a
hierarchical triplet may be decomposed in an inner binary and an outer one.
Indeed, as the separation of the outer binary grows, the CoM of the inner
binary will eventually move out of its NCZ, thus violating the assumptions on
which the derivation of the quadrupole and quadrupole-octupole formulae
relies.

We have described two remedies to this problem. When using the leading order
quadrupole formula, it suffices to express the waveforms in terms of
appropriate conserved quantities (namely the total linear momentum) to
eliminate the observed spurious behavior. When using the quadrupole-octupole
formula, the simplest approach is instead to compute the multipoles in an
inertial frame whose origin instantaneously coincides with the CoM of the
inner binary. Remarkably, neither of these two fixes seems to work for a
system of \textit{four} bodies\footnote{For a system of two relativistic binaries at  sufficiently
large mutual separations, even if we set the origin in the CoM of one binary, the NCZ corresponding to the second binary will
\textit{not} contain the CoM of that binary, hence the ``standard'' PN formalism is not applicable.},
 for which a more sophisticated approach should
be developed.

Finally, we have shown that, if one were to compute the GW emission from a
binary or triple system by integrating directly the equations for the linear
metric perturbations over flat space, one would obtain GW amplitudes that are
wrong by a factor $\sim 2$. We have found that this is related to the fact
that the derivation of the quadrupole formula is quite subtle and actually
requires one to use the \textit{non-linear} Einstein equations.

\section*{Acknowledgments}

The work of MB was supported in part by the ERC Project No. 267117 (DARK)
hosted by Universit\'e Pierre et Marie Curie (UPMC) - Paris 6, PI J. Silk. MB
acknowledge the CINECA award under the ISCRA initiative, for the availability
of high performance computing resources and support. This project has received
funding from the European Union's Horizon 2020 research and innovation
programme under the Marie Sklodowska-Curie grant agreement No 690904. For the
numerical simulations, we have made use of the Horizon Cluster, hosted by the
Institut d'Astrophysique de Paris. We thank Stephane Rouberol for running
smoothly this cluster for us. AS is supported by the Royal Society.

\clearpage
\section*{References}

\bibliographystyle{iopart-num}
\bibliography{bibliography} 

\providecommand{\newblock}{}
\begin{thebibliography}{10}
\expandafter\ifx\csname url\endcsname\relax
  \def\url#1{{\tt #1}}\fi
\expandafter\ifx\csname urlprefix\endcsname\relax\def\urlprefix{URL }\fi
\providecommand{\eprint}[2][]{\href{http://arxiv.org/abs/#2}{arXiv:#2}}

\bibitem{Pretorius2005}
Pretorius F 2005 {\em Phys. Rev. Lett.\/} {\bf 95} 121101
  [\eprint{gr-qc/0507014}]

\bibitem{Campanelli2005}
Campanelli M, Lousto C~O, Marronetti P and Zlochower Y 2006 {\em Phys. Rev.
  Lett.\/} {\bf 96} 111101 [\eprint{gr-qc/0511048}]

\bibitem{Baker2005}
Baker J~G, Centrella J, Choi D~I, Koppitz M and van Meter J 2006 {\em Phys.
  Rev. Lett.\/} {\bf 96} 111102 [\eprint{gr-qc/0511103}]

\bibitem{Alcubierre2008_book}
Alcubierre M 2008 {\em {Introduction to 3+1 Numerical Relativity}\/} (Oxford
  University Press)

\bibitem{Lehner2014}
Lehner L and Pretorius F 2014 {\em Annual Review of Astronomy and
  Astrophysics\/} {\bf 52} 661--694
  [\eprint{https://doi.org/10.1146/annurev-astro-081913-040031}]
  \urlprefix\url{https://doi.org/10.1146/annurev-astro-081913-040031}

\bibitem{Blanchet2014}
Blanchet L 2014 {\em Living Reviews in Relativity\/} {\bf 17} 2 ISSN 1433-8351
  \urlprefix\url{http://dx.doi.org/10.12942/lrr-2014-2}

\bibitem{Poisson2004}
Poisson E 2004 {\em Living Rev. Relativity\/} {\bf 7} 6
  [\eprint{gr-qc/0306052}]
  \urlprefix\url{http://www.livingreviews.org/lrr-2004-6}

\bibitem{Hulse1974}
Hulse R and Taylor J 1975 {\em Astrophys.J.\/} {\bf 195} L51--L53

\bibitem{GW150914}
{Abbott} B~P, {Abbott} R, {Abbott} T~D, {Abernathy} M~R, {Acernese} F, {Ackley}
  K, {Adams} C, {Adams} T, {Addesso} P, {Adhikari} R~X and et~al 2016 {\em
  Physical Review Letters\/} {\bf 116} 061102 [\eprint{1602.03837}]

\bibitem{GW170104}
Abbott B~P {\em et~al.\/} (LIGO Scientific and Virgo Collaboration) 2017 {\em
  Phys. Rev. Lett.\/} {\bf 118}(22) 221101
  \urlprefix\url{https://link.aps.org/doi/10.1103/PhysRevLett.118.221101}

\bibitem{TheLIGOScientific2016pea}
Abbott B~P {\em et~al.\/} (Virgo, LIGO Scientific) 2016 {\em Phys. Rev.\/} {\bf
  X6} 041015 [\eprint{1606.04856}]

\bibitem{Taylor1989}
Taylor J~H and Weisberg J 1989 {\em Astrophys.J.\/} {\bf 345} 434--450

\bibitem{Taylor1982}
Taylor J and Weisberg J 1982 {\em Astrophys.J.\/} {\bf 253} 908--920

\bibitem{Damour1992}
Damour T and Taylor J~H 1992 {\em Phys. Rev. D\/} {\bf 45} 1840--1868

\bibitem{TheLIGOScientific2016src}
Abbott B~P {\em et~al.\/} (LIGO Scientific and Virgo Collaborations) 2016 {\em
  Phys. Rev. Lett.\/} {\bf 116}(22) 221101
  \urlprefix\url{https://link.aps.org/doi/10.1103/PhysRevLett.116.221101}

\bibitem{Schafer1987}
{Sch{\"a}fer} G 1987 {\em Physics Letters A\/} {\bf 123} 336--339

\bibitem{Konigsdorffer2003}
{K{\"o}nigsd{\"o}rffer} C, {Faye} G and {Sch{\"a}fer} G 2003 {\em \prd\/} {\bf
  68} 044004 [\eprint{gr-qc/0305048}]

\bibitem{Lousto2008}
{Lousto} C~O and {Nakano} H 2008 {\em Classical and Quantum Gravity\/} {\bf 25}
  195019 [\eprint{0710.5542}]

\bibitem{Galaviz2011}
{Galaviz} P and {Br{\"u}gmann} B 2011 {\em \prd\/} {\bf 83} 084013
  [\eprint{1012.4423}]

\bibitem{Kozai1962}
{Kozai} Y 1962 {\em \aj\/} {\bf 67} 591

\bibitem{Lidov1962}
{Lidov} M~L 1962 {\em \planss\/} {\bf 9} 719--759

\bibitem{Ford2000}
{Ford} E~B, {Kozinsky} B and {Rasio} F~A 2000 {\em \apj\/} {\bf 535} 385--401

\bibitem{Naoz2016}
{Naoz} S 2016 {\em \araa\/} {\bf 54} 441--489 [\eprint{1601.07175}]

\bibitem{Antonini2016}
{Antonini} F, {Chatterjee} S, {Rodriguez} C~L, {Morscher} M, {Pattabiraman} B,
  {Kalogera} V and {Rasio} F~A 2016 {\em \apj\/} {\bf 816} 65
  [\eprint{1509.05080}]

\bibitem{Antonini2017}
{Antonini} F, {Toonen} S and {Hamers} A~S 2017 {\em \apj\/} {\bf 841} 77
  [\eprint{1703.06614}]

\bibitem{Silsbee2017}
{Silsbee} K and {Tremaine} S 2017 {\em \apj\/} {\bf 836} 39
  [\eprint{1608.07642}]

\bibitem{Blaes2002}
{Blaes} O, {Lee} M~H and {Socrates} A 2002 {\em \apj\/} {\bf 578} 775--786
  [\eprint{astro-ph/0203370}]

\bibitem{Iwasawa2006}
{Iwasawa} M, {Funato} Y and {Makino} J 2006 {\em \apj\/} {\bf 651} 1059--1067
  [\eprint{astro-ph/0511391}]

\bibitem{Iwasawa2008}
{Iwasawa} M, {Funato} Y and {Makino} J 2008 {\em ArXiv e-prints\/}
  [\eprint{0801.0859}]

\bibitem{Hoffman2007}
{Hoffman} L and {Loeb} A 2007 {\em \mnras\/} {\bf 377} 957--976
  [\eprint{astro-ph/0612517}]

\bibitem{Pau2010}
{Amaro-Seoane} P, {Sesana} A, {Hoffman} L, {Benacquista} M, {Eichhorn} C,
  {Makino} J and {Spurzem} R 2010 {\em \mnras\/} {\bf 402} 2308--2320
  [\eprint{0910.1587}]

\bibitem{Haehnelt1994}
{Haehnelt} M~G 1994 {\em \mnras\/} {\bf 269} 199 [\eprint{astro-ph/9405032}]

\bibitem{Jaffe2003}
{Jaffe} A~H and {Backer} D~C 2003 {\em \apj\/} {\bf 583} 616--631
  [\eprint{astro-ph/0210148}]

\bibitem{Wyithe2003}
{Wyithe} J~S~B and {Loeb} A 2003 {\em \apj\/} {\bf 590} 691--706
  [\eprint{astro-ph/0211556}]

\bibitem{Enoki2004}
{Enoki} M, {Inoue} K~T, {Nagashima} M and {Sugiyama} N 2004 {\em \apj\/} {\bf
  615} 19--28 [\eprint{astro-ph/0404389}]

\bibitem{Sesana2004}
{Sesana} A, {Haardt} F, {Madau} P and {Volonteri} M 2004 {\em \apj\/} {\bf 611}
  623--632 [\eprint{astro-ph/0401543}]

\bibitem{Sesana2005}
{Sesana} A, {Haardt} F, {Madau} P and {Volonteri} M 2005 {\em \apj\/} {\bf 623}
  23--30 [\eprint{astro-ph/0409255}]

\bibitem{Jenet2005}
{Jenet} F~A, {Hobbs} G~B, {Lee} K~J and {Manchester} R~N 2005 {\em \apjl\/}
  {\bf 625} L123--L126 [\eprint{astro-ph/0504458}]

\bibitem{Rhook2005}
{Rhook} K~J and {Wyithe} J~S~B 2005 {\em \mnras\/} {\bf 361} 1145--1152
  [\eprint{astro-ph/0503210}]

\bibitem{Barausse2012}
{Barausse} E 2012 {\em \mnras\/} {\bf 423} 2533--2557 [\eprint{1201.5888}]

\bibitem{Klein2016}
{Klein} A, {Barausse} E, {Sesana} A, {Petiteau} A, {Berti} E, {Babak} S, {Gair}
  J, {Aoudia} S, {Hinder} I, {Ohme} F and {Wardell} B 2016 {\em \prd\/} {\bf
  93} 024003 [\eprint{1511.05581}]

\bibitem{Audley2017}
Audley H {\em et~al.\/} 2017  [\eprint{1702.00786}]

\bibitem{Holman1997}
{Holman} M, {Touma} J and {Tremaine} S 1997 {\em \nat\/} {\bf 386} 254--256

\bibitem{Campanelli2006}
{Campanelli} M, {Dettwyler} M, {Hannam} M and {Lousto} C~O 2006 {\em \prd\/}
  {\bf 74} 087503 [\eprint{astro-ph/0509814}]

\bibitem{Campanelli2007}
Campanelli M, Lousto C~O and Zlochower Y 2008 {\em Phys. Rev.\/} {\bf D77}
  101501 [\eprint{0710.0879}]

\bibitem{Lousto2007}
Lousto C~O and Zlochower Y 2008 {\em Phys. Rev.\/} {\bf D77} 024034
  [\eprint{0711.1165}]

\bibitem{Asada2009}
{Asada} H 2009 {\em \prd\/} {\bf 80} 064021 [\eprint{0907.5091}]

\bibitem{Yamada2016}
Yamada K and Asada H 2016 {\em Phys. Rev. D\/} {\bf 93}(8) 084027
  \urlprefix\url{https://link.aps.org/doi/10.1103/PhysRevD.93.084027}

\bibitem{Einstein1918a}
{Einstein} A 1918 {\em Sitzungsberichte der K{\"o}niglich Preu{\ss}ischen
  Akademie der Wissenschaften (Berlin)\/}  154--167
  \urlprefix\url{http://echo.mpiwg-berlin.mpg.de/MPIWG:8HSP60BU}

\bibitem{Landau_book1975}
{Landau} L~D and {Lifshitz} E~M 1975 {\em {The classical theory of fields}\/}
  (Oxford: Pergamon Press)

\bibitem{Thorne1980}
Thorne K~S 1980 {\em Rev. Mod. Phys.\/} {\bf 52} 299--339

\bibitem{Bonetti2016}
{Bonetti} M, {Haardt} F, {Sesana} A and {Barausse} E 2016 {\em \mnras\/} {\bf
  461} 4419--4434 [\eprint{1604.08770}]

\bibitem{Maggiore2007_book}
Maggiore M 2007 {\em {Gravitational Waves. Vol. 1: Theory and Experiments}\/}
  Oxford Master Series in Physics (Oxford University Press) ISBN 9780198570745,
  9780198520740
  \urlprefix\url{http://www.oup.com/uk/catalogue/?ci=9780198570745}

\bibitem{MTW}
{Misner} C~W, {Thorne} K~S and {Wheeler} J~A 1973 {\em {Gravitation}\/} ({San
  Francisco: W.H.~Freeman and Co.})

\bibitem{ADM1959}
{Arnowitt} R, {Deser} S and {Misner} C~W 1959 {\em Physical Review\/} {\bf 116}
  1322--1330

\bibitem{Blanchet2002}
{Blanchet} L 2002 {\em Living Reviews in Relativity\/} {\bf 5} 3
  [\eprint{gr-qc/0202016}]

\bibitem{Press1977}
{Press} W~H 1977 {\em \prd\/} {\bf 15} 965--968

\bibitem{DS85}
{Damour} T and {Sch{\"a}fer} G 1985 {\em General Relativity and Gravitation\/}
  {\bf 17} 879--905

\bibitem{Georg2015}
{Georg} I and {Sch{\"a}fer} G 2015 {\em Classical and Quantum Gravity\/} {\bf
  32} 145001 [\eprint{1503.04618}]

\bibitem{Goldstein1950}
{Goldstein} H 1950 {\em {Classical mechanics}\/}

\bibitem{Pati2000}
{Pati} M~E and {Will} C~M 2000 {\em \prd\/} {\bf 62} 124015
  [\eprint{gr-qc/0007087}]

\bibitem{BDI1995}
Blanchet L, Damour T and Iyer B~R 1995 {\em Phys. Rev. D\/} {\bf 51}(10)
  5360--5386 \urlprefix\url{https://link.aps.org/doi/10.1103/PhysRevD.51.5360}

\bibitem{BDIWW1995}
Blanchet L, Damour T, Iyer B~R, Will C~M and Wiseman A~G 1995 {\em Phys. Rev.
  Lett.\/} {\bf 74}(18) 3515--3518
  \urlprefix\url{https://link.aps.org/doi/10.1103/PhysRevLett.74.3515}

\bibitem{BD86}
Blanchet L and Damour T 1986 {\em Phil. Trans. Roy. Soc. Lond. A\/} {\bf 320}
  379--430
  \urlprefix\url{http://rsta.royalsocietypublishing.org/content/320/1555/379}

\bibitem{BD89}
Blanchet L and Damour T 1989 {\em Annales Inst. H. Poincar{\'e} Phys.
  Th\'eor.\/} {\bf 50} 377--408
  \urlprefix\url{http://www.numdam.org/item/AIHPA_1989__50_4_377_0}

\bibitem{B98mult}
Blanchet L 1998 {\em Class. Quant. Grav.\/} {\bf 15} 1971--1999
  [\eprint{gr-qc/9801101}]
  \urlprefix\url{http://stacks.iop.org/0264-9381/15/i=7/a=013}

\bibitem{PB02}
Poujade O and Blanchet L 2002 {\em Phys. Rev. D\/} {\bf 65}(12) 124020
  \urlprefix\url{https://link.aps.org/doi/10.1103/PhysRevD.65.124020}

\bibitem{Weinberg1972book}
{Weinberg} S 1972 {\em {Gravitation and Cosmology: Principles and Applications
  of the General Theory of Relativity}\/}

\bibitem{Poisson2014_book}
{Poisson} E and {Will} C~M 2014 {\em {Gravity}\/}

\end{thebibliography}

\clearpage

\appendix

\section{Definitions}
\label{sec:app_A}
In this appendix we summarize the explicit leading order expressions for the
quadrupole and octupole mass radiation, as well as for the current quadrupole
radiation.

The second and third mass moments are defined from the time-time component of
the matter stress-energy tensor as
%
\begin{eqnarray}
M^{ij}(t) = \int \ud^3\bm{x} \ c^{-2} T^{00}(t,\bm{x}) x^{\langle i} x^{j \rangle},\nonumber\\
M^{ijk}(t) = \int \ud^3\bm{x} \ c^{-2} T^{00}(t,\bm{x}) x^{\langle i} x^j x^{k
  \rangle}\, ,
\end{eqnarray}
where $\langle \rangle$ represents the symmetric trace-free (STF) operator, i.e., (see~\cite{Maggiore2007_book} \S 3.5.1)
%
\begin{equation}
\fl x^{\langle i} x^{j \rangle} = x^i x^j - \frac{1}{3} \delta^{ij}r^2\, ,
~
x^{\langle i} x^j x^{k \rangle} = x^i x^j x^k - \frac{1}{5}
  \left(\delta^{ij}r^2 x^k + \delta^{ik}r^2 x^j + \delta^{jk}r^2 x^i\right)\, .
\end{equation}
The current quadrupole moment is defined by
%
\begin{equation}\label{eq:curr_quad_mom}
S^{ijk} = \int \ud^3\bm{x} \ [ x^i j^{kj} + x^j j^{ki}]\, ,
\end{equation}
where 
%
\begin{equation}
j^{ij} = \frac{1}{c}\left(x^i T^{0j} - x^j T^{0i}\right)
\end{equation}
is the angular momentum density tensor, which is connected to the angular
momentum density vector by $j^{ij} = \varepsilon^{ij}_{~~k}j^k$. Alternatively, one
may define the STF quadrupole tensor
%
\begin{equation}
J^{ij} = \mathop{\rm STF}_{ij}\varepsilon_{abi} \int \ud^3\bm{x} \ x^{\langle j}
x^{a\rangle} T^{0b}\, ,
\end{equation}
and replace $S^{abk}$ by $-2 \varepsilon_{~l}^{k~a} J^{bl}$ in the waveform,
using the fact that $S^{abk} n_k$ and $-2 \varepsilon_{~l}^{k~a} J^{bl} n_k$
have the same transverse trace-free part (with respect to $\bm{n}$). The
advantage of working with $J^{ij}$ is that it belongs to an irreducible
representation of SO(3).

The expression of the GW waveform up to the next-to-leading order in the
TT gauge is finally given by
%
\begin{eqnarray}
\fl h_{ij}^{\rm TT}(t,\bm{x}) 
\approx \frac{2 G}{R c^4} \Lambda_{ijkl}(\bm{n}) & \Biggl[ \frac{\ud^2
M^{kl}(t_{\rm ret})}{\ud t^2} \nonumber\\
\fl & 
+ \frac{n_m}{3c} \frac{\ud^3 M^{klm}(t_{\rm ret})}{\ud t^3} + \frac{2 n_m}{3c} 
\frac{\ud^2 S^{klm}(t_{\rm ret})}{dt^2} \Biggr]+
      \mathcal{O}\Big(\frac{1}{c^6}\Big)\, ,
\end{eqnarray}
where the projector tensor $\Lambda_{ijkl}(\bm{n})$ is defined in terms of the
GW propagation direction $\bm{n}$ (see~\cite{MTW} \S 36.10 and Box 35.1):
%
\begin{equation}
\Lambda_{ijkl}(\bm{n}) = \mathcal{P}_{ik}\mathcal{P}_{jl} - 
\frac{1}{2}\mathcal{P}_{ij}\mathcal{P}_{kl}, \quad \mathcal{P}_{ij} =
\delta_{ij} - n_i n_j\, .
\end{equation}

\section{Calculations}
\label{sec:app_C}

In this appendix, we perform the explicit calculation of
%
\begin{equation}
\label{eq:H_generic}
\fl \bar{h}^{ij} (t,\bm{x}) = \frac{4G}{c^4}\int 
\frac{\ud^3\bm{x}'}{|\bm{x}-\bm{x}'|}\biggl[ T^{ij}
+ \frac{1}{4\pi G} \left(\partial^i\phi \, \partial^j\phi-
\frac{1}{2}\delta^{ij}\partial_k\phi \, \partial^k\phi\right)\biggr] + 
\mathcal{O}\Big(\frac{1}{c^5}\Big)\, .
\end{equation} 
Since the first term of eq.~\ref{eq:H_generic} is trivial to evaluate because
it has a compact support, we focus on the second one, where one is not a priori 
allowed to approximate $|\bm{x}-\bm{x}'|$ by $R$ under the integral, since the
integration extends up to spatial infinity. We perform the calculation by
considering only the term $\partial^i\phi \, \partial^j\phi$. Indeed,
$\delta^{ij}\partial_k\phi \, \partial^k\phi$, which is simply the trace of
$\partial^i\phi\, \partial^j\phi$ multiplied by a Kronecker delta, disappears
when taking the TT projection.

Inserting the expression for the Newtonian gravitational potential, we find
%
\begin{eqnarray}
\label{eq:off_diag}
\fl \frac{1}{4\pi G}\int
  \frac{\ud^3\bm{x}'}{|\bm{x}-\bm{x}'|} \partial^i\phi\, \partial^j\phi 
&=\frac{G}{4\pi}\int 
\frac{\ud^3\bm{x}' \ud^3\bm{y}' \ud^3\bm{y}''}{|\bm{x}-\bm{x}'|}
\rho(\bm{y}')\rho(\bm{y}'') \times\nonumber\\
&\times \frac{\partial}{\partial x'^{i}}
\left(\frac{1}{|\bm{x}'-\bm{y}'|}\right)  
\frac{\partial}{\partial x'^{j}}\left(\frac{1}{|\bm{x}'-\bm{y}''|}\right)\, .
\end{eqnarray}
After transforming the derivative $\partial/\partial x'^i$ that acts on
$|\bm{x}'-\bm{y}'|^{-1}$ into $-\partial/\partial y'^i$ by virtue of the
translation invariance of $\bm{x}'-\bm{y}'$, and similarly for
$\partial/\partial x'^j$, we may change the order of integration, so
that $\partial^2/(\partial y'^i \partial y''^j)$ can be put outside the
integral with respect to $\bm{x}'$. With this trick,
eq.~\ref{eq:off_diag} becomes
%
\begin{equation}
  \fl  \frac{1}{4\pi G}\int
  \frac{\ud^3\bm{x}'}{|\bm{x}-\bm{x}'|} \partial^i\phi\, \partial^j\phi = G \int
  \ud^3\bm{y}'   \ud^3\bm{y}''
  \rho(\bm{y}')\rho(\bm{y}'') \frac{\partial^2}{\partial
    y'^{i}\partial y''^{j}} g(\bm{x}',\bm{y}',\bm{y}'')\,, 
\end{equation}
where $g$ satisfies the Poisson equation
$\Delta g(\bm{x},\bm{y}',\bm{y}'') = |\bm{x}-\bm{y}'|^{-1}
|\bm{x}-\bm{y}''|^{-1}$
in the sense of distributions. It is straightforward to check that the relevant
solution is 
$g=\ln (|\bm{x}-\bm{y}'|+|\bm{x}-\bm{y}''|+|\bm{y}'-\bm{y}''|)+{\rm constant}$
(see, e.g.,
p.355 in \cite{Poisson2014_book}), from which one infers the asymptotic behavior
%
\begin{equation}
\frac{\partial^2}{\partial y^{'i}\partial y^{''j}}K = 
-\frac{\hat{n}^i \hat{n}^j-\delta^{ij}}{2 R |\bm{y}'-\bm{y}''|} +
\mathcal{O}\Big(\frac{1}{R^2}\Big)\, .
\end{equation}
At large distance $R$ from the origin, eq.~\ref{eq:H_generic} then reduces to
%
\begin{eqnarray}
\fl \bar{h}^{ij} = \frac{4G}{c^4 R}\int \ud^3\bm{x}' T^{ij}
- \frac{2 G^2}{R c^4}\int \ud^3\bm{y}'
  \ud^3\bm{y}''\rho(\bm{y}')\rho(\bm{y}'') \frac{\hat{n}^i
  \hat{n}^j}{|\bm{y}'-\bm{y}''|} + \mathcal{O}\Big(\frac{1}{R^2 c^5}\Big) \,.
\end{eqnarray}


\end{document}